\documentclass[%
 reprint,
prl,
nofootinbib]{revtex4-2}

\usepackage{dblfloatfix} 
\usepackage{mathtools}
\usepackage{amsmath, amssymb}
\usepackage{blindtext}
\usepackage{lipsum}
\usepackage{graphicx}
\usepackage{dcolumn}
\usepackage{bm}
\usepackage{amsmath,xparse}

\usepackage[resetlabels,labeled]{multibib}
\newcites{S}{Supplementary References}

\NewDocumentCommand{\prebig}{mme{_^}}{%
  {\mathclose{\vphantom{#1#2}}\leftscripts{#3}{#4}}#1#2%
}

\ExplSyntaxOn
\NewDocumentCommand{\leftscripts}{mm}
 {
  \IfNoValueTF{#1}
   {
    \IfValueT{#2}{\sp{#2}}
   }
   {
    \IfNoValueTF{#2}
     {
      \sb{#1}
     }
     {
      \mammoth_align_scripts:nn { #1 } { #2 }
     }
   }
 }
\cs_new_protected:Nn \mammoth_align_scripts:nn
 {
  \hbox_set:Nn \l_tmpa_box { \hfil$\scriptstyle#1$ }
  \hbox_set:Nn \l_tmpb_box { \hfil$\scriptstyle#2$ }
  \dim_set:Nn \l_tmpa_dim
   {
    \dim_max:nn { \box_wd:N \l_tmpa_box } { \box_wd:N \l_tmpb_box }
   }
  \sb{ \hbox_to_wd:nn { \l_tmpa_dim } { \hbox_unpack:N \l_tmpa_box } }
  \sp{ \hbox_to_wd:nn { \l_tmpa_dim } { \hbox_unpack:N \l_tmpb_box } }
 }
\ExplSyntaxOff

\begin{document}

\preprint{APS/123-QED}

\title{ \emph{Driven} spin dynamics enhances cryptochrome magnetoreception: \\Towards live quantum sensing} 


\author{Luke D.\ Smith}
\author{Farhan T.\ Chowdhury}
\author{Iona Peasgood}
\author{Nahnsu Dawkins}%
\author{Daniel R.\ Kattnig}
\email{Corresponding author: \\ d.r.kattnig@exeter.ac.uk}
 \affiliation{%
		Living Systems Institute and Department of Physics \\ University of Exeter, Stocker Road, Exeter EX4 4QD, United Kingdom}
\date{\today}

\begin{abstract}
The mechanism underlying magnetoreception has long eluded explanation. A popular hypothesis attributes this sense to the quantum coherent spin dynamics of spin-selective recombination reactions of radical pairs in the protein cryptochrome. However, concerns about the validity of the hypothesis have been raised as unavoidable inter-radical interactions, such as strong electron-electron dipolar coupling, appear to suppress its sensitivity. We demonstrate that this can be overcome by driving the spin system through a modulation of the inter-radical distance. It is shown that this dynamical process markedly enhances geomagnetic field sensitivity in strongly coupled radical pairs via a Landau-Zener type transition between singlet and triplet states. These findings suggest that a ``live" harmonically driven magnetoreceptor can be more sensitive than its ``dead" static counterpart.
 
\end{abstract}

\maketitle
\textit{Introduction}.---The geomagnetic field provides a frame of reference that living systems use towards essential functions \cite{Mouritsen2001, Johnsen2005, Johnsen2008, Kominis2015, Mouritsen2018}. Migratory birds exemplify this in their reliance on an internal compass that aids their navigation to breeding and wintering sites \cite{Mouritsen2015, Wiltschko2019}. Although what underlies this fine-tuned compass sense remains unsettled \cite{Nordmann2017}, growing evidence suggests its reliance on magnetosensitivity acquired through the quantum spin dynamics of a radical pair recombination reaction mediated by the blue-light sensitive flavoprotein \emph{cryptochrome} \cite{Ball2011, Lambert2013, Mohseni2014, Marais2018, Kim2021}.

The widely studied radical pair mechanism (RPM) involves the quantum dynamics of two spatially separated unpaired electrons \cite{Schulten1978a, Ritz2000, Hore2016}. Their combined spin angular momentum can be described in terms of singlet/triplet states, which are measurable insofar as they are subjected to distinct recombination reactions giving rise to different chemical products. Consequently, magnetosensitivity is elicited as a result of coherent singlet-triplet interconversion occurring predominantly due to hyperfine couplings of electron spins with surrounding magnetic nuclei and their interaction with an applied magnetic field. This basic mechanism forms the cornerstone of the quantum compass hypothesis, providing tentative explanations for many of the traits of avian magnetoreception. Theoretical studies have provided deeper insight, such as the inter-relation of coherence and magnetosensitivity in both toy models \cite{Gauger2011, Hogben2012, Cai2013, Zhang2014, Carrillo2015, Le2020, Kominis2020, Jain2021} and more realistically complex ones \cite{Cai2010, Lee2014, Hiscock2016,  Atkins2019, Fay2019, Smith2022}. In cryptochrome, a commonly adopted model supported by \textit{in vitro} studies assumes that a photo-induced electron transfer forms a radical pair between a flavin anion radical (FAD$^{\bullet-}$) and a tryptophan radical cation (TrpH$^{\bullet+}$) \cite{Kattnig2016, Kerpal2019, Xu2021}. Alternative reaction mechanisms, such as dark-state oxidation schemes, have also been proposed and investigated \cite{Wiltschko2016, Hammad2020}. 

However, many facets, including the identity of the radical pair relevant \textit{in vivo}, remain open problems. Studies of magnetosensitivity in more realistic settings, where the presence of many hyperfine couplings and environmental noise can constrain coherence lifetimes to a few microseconds \cite{Maeda2012, Kattnig2016,Kattnig2016a, Atkins2019, Kobylkov2019}, are limited. Furthermore, inter-radical couplings, such as electron-electron dipolar (EED) and exchange interactions, can suppress magnetosensitivity \cite{ODea2005, Hiscock2016a, Hiscock2017, Babcock2020}, but were neglected in a majority of theoretical studies. While exchange interaction was shown to be negligible for selected cryptochromes via time resolved EPR spectroscopy \cite{Schiemann2007, Nohr2017}, due to the close vicinity of the recombining radical centres  (approximately $1.5$\,nm) EED coupling is unavoidable \cite{Hiscock2016a, Babcock2020}. To resolve this, a mutual compensation of exchange and EED interactions was suggested \cite{Efimova2008}, but it was ineffective for radical pairs involving the flavin radical \cite{Babcock2020}. The quantum Zeno effect could partly alleviate the detrimental effect due to inter-radical coupling \cite{Dellis2012}, but this requires fast triplet recombination which is unlikely in cryptochrome.  Three-radical models \cite{Kattnig2017, Kattnig2017a} show enhanced magnetosensitivity in the presence of EED coupling by suitable placement of an inert radical bystander \cite{Keens2018,Babcock2020}, or by postulating a spin-selective recombination reaction involving a radical of the primary pair and a scavenger radical \cite{Babcock2021}. While these models demonstrate effective magnetosensitivity, the resulting reaction schemes are complex and currently lack directly supporting evidence.

It has been shown that environmental fluctuations can potentially counteract the detrimental effect of inter-radical couplings \cite{Kattnig2016b}, suggesting further enhancements may be possible from structured molecular dynamics. In this work, we consider a model that compliments and extends on the established RPM by modulating inter-radical distance as a function of time. This aims to incorporate driving contributions of a ``live" magnetoreceptor arising from, or actively sustained by, protein dynamics at physiological temperatures \cite{Kattnig2018}. By approximating driving as harmonic motion that modulates both exchange/EED interactions and recombination rate, we find that sensitivity can be vastly amplified.

\textit{Driven model of magnetoreception.---}
In our model, radical pairs undergo coherent evolution subject to time-dependent harmonic driving, with the Hamiltonian $\hat{H}(t)$ comprising Zeeman, hyperfine, and time-dependent exchange and EED
interactions. The radical pair reaction of A$^{\bullet-}$ and B$^{\bullet+}$ involves singlet $^{1}$[A$^{\bullet-}$/B$^{\bullet+}$] and triplet $^{3}$[A$^{\bullet-}$/B$^{\bullet+}$] interconversion, recombination with rate $k_{b}(t)$, and forward reaction with rate $k_{f}$ as follows
\begin{center}
\includegraphics[]{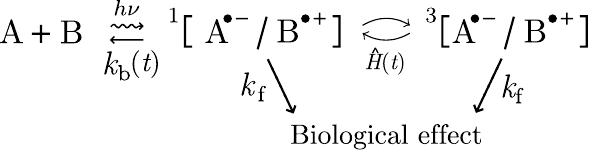}
\vspace{-0.6em}
\end{center}
The initial singlet state of this reaction is given by a spin density operator $\hat{\rho}(0) = \frac{\hat{P}_{S}}{Z},$ where $\hat{P}_{S}$ is the singlet projection operator and $Z=Z_{A}Z_{B}$ denotes dimension of the nuclear subspace associated with the two radicals. Time evolution of $\hat{\rho}(t)$ is described by the master equation
\begin{align}
    \frac{\mathrm{d}{\hat{\rho}}(t)}{\mathrm{d}t} = -i[\hat{H}(t), \hat{\rho}(t)] - \frac{k_{b}(t)}{2}\{\hat{P}_{S}, \hat{\rho}(t)\} - k_{f}\hat{\rho}(t), \label{eq.master_equation}
\end{align}
where $[\,]$ represents the commutator, and $\{\}$ the anticommutator. The solution to Eq.~(\ref{eq.master_equation}) is $\hat{\rho}(t)=\hat{U}(t,0) \hat{\rho}(0) \hat{U}^{\dagger}(t,0)$, 
where the time evolution operator is
\begin{align}
    \hat{U}(t,0) = \mathcal{T} \mathrm{exp}\left[-i\int_{0}^{t}\hat{H}_{\mathrm{eff}}(\tau)\mathrm{d}\tau\right],
\end{align}
with an effective Hamiltonian given by
\begin{align}
    \hat{H}_{\mathrm{eff}}(t) = \hat{H}(t) - i\left( \frac{k_{b}(t)}{2}\hat{P}_{S} + \frac{k_{f}}{2}\hat{I}\right).
\end{align}
The time-dependent recombination yield due to the singlet channel is found using
\begin{align}
    \Phi = \int_{0}^{\infty}k_{b}(t)p_{S}(t)\mathrm{d}t, \label{eq.singlet_yield}
\end{align}
where $p_{S}(t) = \mathrm{Tr}[\hat{P}_{S}\hat{\rho}(t)] =\frac{1}{Z}\sum_{i} \langle \psi_{i}(t) \vert \hat{P}_{S} \vert \psi_{i}(t) \rangle$ and the singlet recombination yield is evaluated by propagating wavefunctions using 
$\vert \psi_{i}(t) \rangle = \hat{U}(t,0) \vert \psi_{i}(0) \rangle$, summed over all $i$ initial singlet states of various nuclear spin configuration. As the effective Hamiltonian is periodic in time, i.e.\ $\hat{H}_{\mathrm{eff}}(t) = \hat{H}_{\mathrm{eff}}(t+T)$ with $T = \nu_{\mathrm{d}}^{-1}$ denoting its period, we utilize Floquet theory to speed up the relevant computations for large driving frequencies $k_{f}, k_{b} \ll \nu_{\mathrm{d}}$ (see Supplemental Material (SM) for more details).

To exemplify key features of a driven radical pair system, we first focus on a simple model comprising a single hyperfine-coupled nitrogen atom ($\hat{I} = 1$) in one radical and no hyperfine interactions in the other. Inter-radical interactions are considered in the form of a scalar coupling $J(r)$ formally corresponding to the exchange interaction but qualitatively also encompassing unavoidable EED coupling. For inter-radical distance modulated as
\begin{align}
r(t) = \frac{\Delta_{\mathrm{d}}}{2}[1-\cos(2\pi \nu_{\mathrm{d}}t)] + r_0,
\end{align}
we choose a singlet recombination rate of the form \cite{Steiner1989}
\begin{align}
k_{b}(t) = k_{b_{0}} \exp[-\beta (r(t)-r_0)],
\end{align}
with $r_0$ specifying the inter-radical distance of the static radical pair, $k_{b_{0}}=2\,\mu$s$^{-1}$, $k_{f}=1\,\mu$s$^{-1}$, $\beta=1.4$\,\AA$^{-1}$ \cite{Moser1992}. Exchange interaction is taken in the same functional form
\begin{align}
J(t) = J_{0} \exp[-\beta (r(t)-r_0)]. 
\end{align}
The single non-zero hyperfine interaction was assumed to be axial, with principal components given by $A_{xx} = A_{yy} = A_{\perp} = -2.6$\,MHz and $A_{zz} = A_{\parallel} = 49.2$\,MHz, representative of the dominating nitrogen atom (N5) in flavin radicals. To assess directional magnetic field effect (MFE), we compute the relative anisotropy 
\begin{align}
\chi = \vert \Phi_{\parallel} - \Phi_{\perp} \vert / \max(\Phi_{\parallel}, \Phi_{\perp}),   
\end{align}
where $\Phi_{\parallel}$ and $\Phi_{\perp}$ are the singlet recombination yields, calculated via Eq.~(\ref{eq.singlet_yield}), for the static magnetic field pointing in parallel and perpendicular directions, respectively. 

\begin{figure}
\centering
	\includegraphics[]{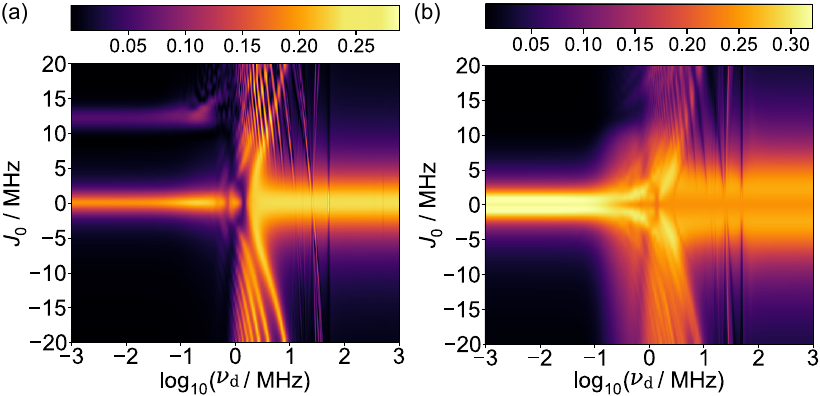}
	\caption{Color maps of the driven radical pair model with EED interaction neglected for a variation of exchange interaction strength $J_{0}$ against driving frequency $\nu_{\mathrm{d}}$. (a) Relative anisotropy $\chi$. (b) Relative entropy of coherence $C_{r}$ evaluated in the singlet-triplet basis. Values of $J_{0}\gtrsim 1\,$MHz create a suppressive effect in the static case, which is removed by including driving in the approximate range of $0 < \nu_{\mathrm{d}} < 10\,$MHz.    \label{Fig.simple_model}}
\vspace{-2.2em}
\end{figure}

Figure \ref{Fig.simple_model}(a) shows relative anisotropy for a variation of exchange interaction strength $J_{0}$ against driving frequency $\nu_{\mathrm{d}}$ and $\Delta_{\mathrm{d}} = 3\,$\AA. For the static case, inter-radical coupling is seen to suppress magnetic field sensitivity for values of $|J_{0}|\gtrsim 1\,$MHz. However, as driving frequency is increased, this suppression is lifted. In particular, a driving frequency in the approximate range of $1$ to $10\,$MHz provides recovery of a MFE for $-20\leq J_{0}\leq20\,$MHz (but extends to even larger values of $|J_{0}|$; see SM). In Fig.~\ref{Fig.simple_model}(b), we take a time-integrated average of the commonly used relative entropy of coherence \cite{Baumgratz2014}, defined as
\begin{align}
\mathcal{C}_{r}[\hat{\sigma}]=S[\mathbb{I}\mathbb{C}(\hat{\sigma})] - S[\hat{\sigma}], 
\end{align}
over parallel and perpendicular directions of the magnetic field with the normalized density operator $\hat{\sigma}$ chosen with respect to the electronic singlet-triplet basis $\{\vert n \rangle \}_{n=1}^{d}$ of the $d$-dimensional Hilbert space. Here, the dephasing operation is given by $\mathbb{IC}(\hat{\sigma}) = \sum_{n} \vert n \rangle \langle n \vert \hat{\sigma} \vert n \rangle \langle n \vert$ and the von Neumann entropy is denoted by $S[\hat{\sigma}]= - \mathrm{Tr}[\hat{\sigma}\log(\hat{\sigma})]$. The broad agreement between Fig.~\ref{Fig.simple_model}(a) and Fig.~\ref{Fig.simple_model}(b) shows that a reinstatement of the MFE is accompanied by a stimulation of coherence, but this relation depends on synchronization with singlet-triplet oscillations of the driven system. We also considered alternative coherence and entanglement measures, with data available in the SM, to provide further explanation of specific features.

To elucidate the physical basis of the observed enhancements, we consider the case of the one-nitrogen radical pair with $A_{\perp} = 0$. With this simplification, the effective Hamiltonian is reducible with blocks labelled by the magnetic quantum number of the nuclear spin $m_{I} \in \{1, 0, -1\}$. Using basis states $\vert T_{+} \rangle$, $\vert T_{0} \rangle$, $\vert T_{-} \rangle$, and $\vert S \rangle$ associated with the triplet and singlet states, respectively, the system Hamiltonian for the magnetic field pointing along the perpendicular axis takes the form
\begin{align}
\hat{H}_{\perp} =
\begin{pmatrix}
am_{I} - J & b & 0 & 0\\
b & -J & b & am_{I}\\
0 & b & -am_{I} - J & 0\\
0 & am_{I} & 0 & J
\end{pmatrix},
\end{align}
where $a=A_{\parallel}/2$, $b=\omega_{0}/\sqrt{2}$, and a similar expression applies for the parallel orientation (see SM). This scenario resembles a Landau-Zener transition \cite{Tully2012, Nelson2020}, where the relevant states of the avoided crossing are $\vert S \rangle$ and $\vert T_{0} \rangle$ that have a constant coupling via the hyperfine interaction for $m_{I} \neq 0$. If inter-radical coupling is static and large, $|J|>>A_{\parallel}$, then the energy separation of $\vert S \rangle$ and $\vert T_{0} \rangle$ traps the system in the singlet state. Driving this system through a modulation of the inter-radical distance introduces a time dependent $J(t)$, periodically decreasing the energy difference and inducing non-adiabatic transitions from $\vert S \rangle$ to $\vert T_{0} \rangle$. Consequently, the trapped singlet population is released which, in the perpendicular orientation, allows evolution to occur between $\vert T_{\pm} \rangle$ states mediated by the magnetic field and without suppression from exchange interaction. This is due to the common energy shift of the triplet states under exchange coupling. Thus the overall process restores the MFE for large $J$. 

For the parallel direction, a similar structure ensues as well, however, the field-dependent interconversion of the triplet states is absent. If the inter-radical coupling is static and moderate, $|J| \sim A_{\parallel}$, the system dynamics is characterized by fast $\vert S \rangle$-$\vert T_0 \rangle$ interconversion. Nevertheless, a periodic reduction of $J$ enables population redistribution of $\vert T_{\pm} \rangle$ states for the perpendicular field direction, which reinstates magnetosensitivity and explains the increase in coherence observed in Fig.~\ref{Fig.simple_model}(b). The basis of this principle extends to systems with more hyperfine couplings and $A_{\perp} \neq 0$, where several avoided-crossings ensue, potentially inducing non-adiabatic transitions. Further analysis on simple models may be found in the SM, where we show that general features persist for a further simplified two-level system, and for the case where the time-dependent recombination rate constant is substituted by its temporal average.

The driving in natural systems may be constrained, such as in charge separation initiated structural rearrangements in a protein leading to radical pair generation \cite{Worster2017, Xu2021}. To address this, we analyzed a model using damped oscillations and found that the MFE may be even further enhanced (Fig.~ S9), suggesting the importance of driving in the early time spin dynamics for freeing the trapped singlet population through avoided crossings. Subsequently a decay in oscillation amplitude aids in enabling efficient recombination at suitable time periods. Additionally, we have used an optimal quantum control approach to maximize the directional sensitivity of the system. For exchange coupling $J_0 = 10$\,MHz, optimizing the absolute anisotropy gives enhancements up to $\chi = 0.73$, exceeding effects realizable through harmonic driving by a factor of 3 (see Fig.~S12).
\begin{figure}[b]
\vspace{-2.0em}
\centering
	\includegraphics[]{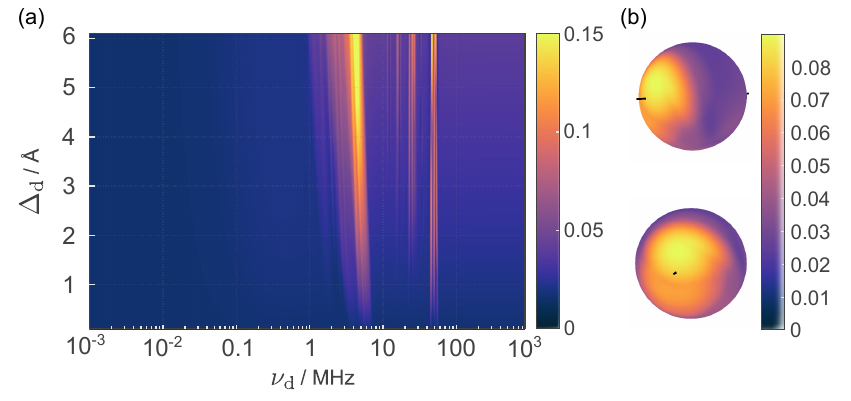}
	\caption{Driven radical pair model with EED interactions included and $J_{0}=0$. (a) Color map of the relative anisotropy, $\chi$ is shown for a variation of oscillation amplitude $\Delta_{\mathrm{d}}$ against driving frequency $\nu_{\mathrm{d}}$. The displacement is assumed to happen along the inter-radical axis. Relative anisotropy is suppressed in the static case, but can be restored for driving frequencies in the approximate range of $1\leq \nu_{\mathrm{d}} \leq 100\,$MHz. (b) Orientation dependence is displayed for an oscillation amplitude of $\Delta_{\mathrm{d}}=2\,$\AA, and $\nu_{\mathrm{d}}=4.4\,$MHz, demonstrating effectiveness with oscillations broadly along the inter-radical axis.  \label{Fig.EED_only}}
	\vspace{-0.3em}
\end{figure}

Thus far a restoration of magnetic field sensitivity, due to driving, has been observed for a model system with axial symmetry subject only to the scalar exchange interaction. It could be argued that the observed enhancement is only due to the intermittent reduction of $J_{0}$, which has a strong dependence on the distance and could be nullified by relatively small-amplitude oscillations. In contrast, the EED interaction decays slowly with respect to $r$ $(\propto 1/r^{3}$) and could be more detrimental to magnetosensitivity. Therefore, we analyze the system again with EED interaction included, using parameters reflecting the relative orientation of the FAD and TrpH radicals in cryptochrome. As this system lacks axial symmetry, the MFE was assessed by the established measure $\Gamma = (\Phi_{max} - \Phi_{min})/\overline\Phi$, 
where $\Phi_{min}$ and $\Phi_{max}$ are the minimum and maximum yield and $\overline{\Phi}$ denotes the average over orientations of the applied magnetic field relative to the spin system (assessed for 2562 orientations).

For $J_{0}=0$, the case without additional exchange coupling, Fig.~\ref{Fig.EED_only}(a) demonstrates that magnetosensitivity can be restored for a range of oscillation amplitudes $\Delta_{\mathrm{d}}$ at an appropriate driving frequency, although the effect is more striking for larger amplitudes. We also observe a small enhancement in magnetosensitivity for oscillations that reduce the inter-radical distance (Fig.~S11). Consequently, the MFE persists as long as there is some driving that creates a time-dependent relative decrease in the EED in tandem with the change in recombination rate. In Fig.~\ref{Fig.EED_only}(b) dependence of the effect on oscillation direction relative to inter-radical axis is analyzed with $\Delta_{\mathrm{d}}=2$\,\AA. We observe that an exact alignment is not required and oscillations that broadly increase the distance are effective. Oscillation amplitudes of 4\,\AA \ and 6\,\AA \ have been considered (Fig.~S10) which support these findings, showing that for increased oscillation amplitude the optimal oscillation direction is one that increases the inter-radical distance and is close to the inter-radical axis. \

\begin{figure}[t]
\centering
	\includegraphics[]{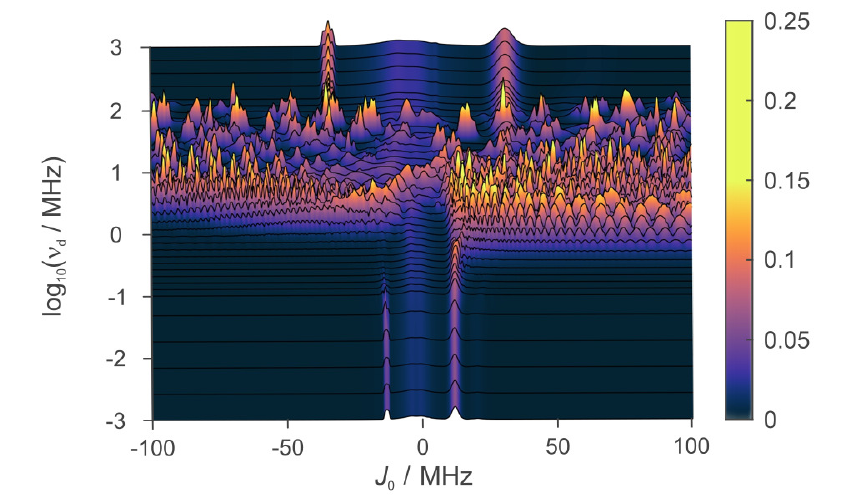}
	\caption{Color map for anisotropy of a driven model with both exchange and EED interactions, shown for variation of driving frequency $\nu_{\mathrm{d}}$ against exchange interaction $J_{0}$. The displacement is assumed to happen along the inter-radical axis. Restorations of anisotropy are observed for driving frequencies in the approximate range of $1 \leq \nu_{\mathrm{d}} \leq 100\,$MHz, even for large exchange interaction strengths in the range $-100\leq J_{0} \leq 100\,$MHz.   \label{Fig.EED_and_exchange}}
\vspace{-2.2em}
\end{figure}

To analyze whether a MFE is possible in driven systems that are subject to both exchange and EED interactions, we have used the above model with exchange and EED included. In Fig.~\ref{Fig.EED_and_exchange} we show the results of this model for a variation of $J_{0}$ in the range of $ -100 \leq J_{0} \leq 100\,$MHz and a choice of a relatively small oscillation amplitude of only $\Delta_{\mathrm{d}}=2$\,\AA. In the static case, we observe two peaks of magnetosensitivity as a function of $J_{0}$. As the system is driven in the range of $1$-$100\,$MHz, complex patterns emerge with markedly enhanced MFEs and a remarkable resilience to large $J_{0}$. Within this range, there is a repetitive, though not strictly periodic, dependence on $J_{0}$. As the driving frequencies are increased beyond $100\,$MHz, the response eventually resembles that of the static case, but with peaks occurring at larger $J_{0}$ such that the time-averaged $J_{0}$ corresponds to the quasi-static $J_{0}$.

\begin{figure}[b]
\centering
	\vspace{-0.5em}
	\includegraphics[]{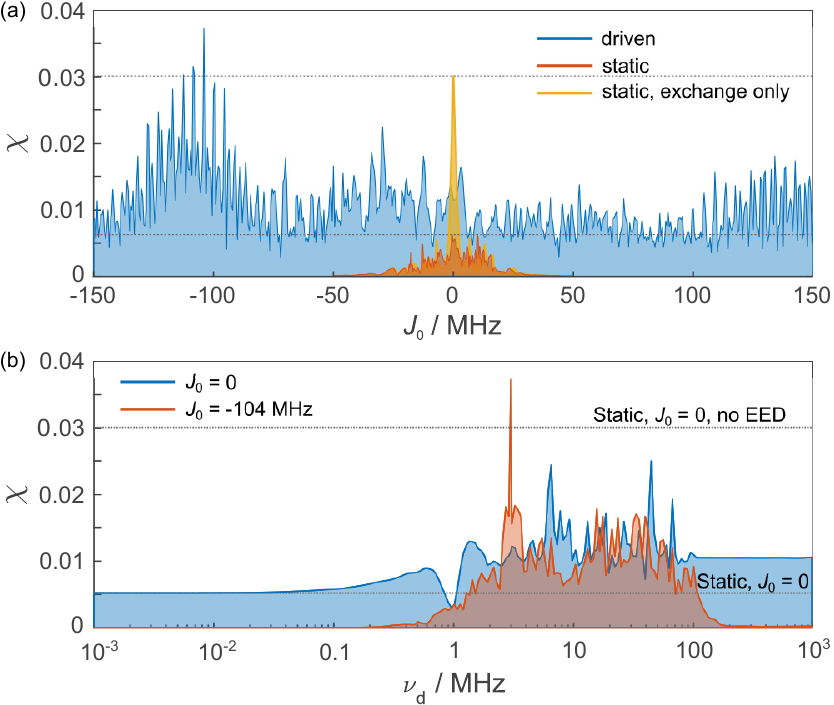}
	\caption{Relative anisotropy $\chi$ of models comprising $4$ nuclear spins against (a) $J_{0}$ for the driven system with EED interaction and static case including or neglecting it, and (b)  $\nu_{\mathrm{d}}$ with $J_{0} = 0\,$MHz or $J_{0}=-104\,$MHz. Dotted lines represent the maximum value of static models shown in (a) at $J_{0} = 0\,$MHz.   \label{Fig.toy_system}}
	\vspace{-1.29em}
\end{figure}

\textit{Magnetic field effect in larger driven systems.---} Since an enhancement in magnetosensitivity for simple driven radical pair systems comprising a single nuclear spin is observed, we now address if this MFE persists as the system is made more complex by extending it to comprise $4$ nuclear spins. Specifically, we consider a flavin-tryptophan radical pair with the established driven model as before, but now include two nuclear spins for each radical with hyperfine couplings corresponding to N5 and N10, and N1 and H1, respectively. We arbitrarily choose a driving frequency of $3\,$MHz, found to be in the effective range for the simpler model, and an amplitude of $2$\,\AA. Results for this model are displayed in Fig.~\ref{Fig.toy_system}(a), where the system in the presence of the time-dependent $J(t)$ is compared to the static scenario for both EED and exchange interactions included, and with the hypothetical scenario with only exchange interaction.

With EED, the data resembles that of the simpler models with enhancements up to a factor of approximately 6 as compared to the static model. For an optimal $J_{0}$, the driven scenario, including both inter-radical interactions, is able to exceed the idealized static scenario for which the EED interaction is plainly neglected. As in the simpler models investigated, the MFE observed is resilient even to large exchange couplings with enhancements persisting in the range of $-150 \leq J_{0} \leq 150\,$MHz.\
In Fig.~\ref{Fig.toy_system}(b) the magnetosensitivity is displayed as a function of the driving frequency for chosen values of $J_{0}$, namely $J_{0}=0$ and $J_{0}=-104$MHz, as a function of the driving frequency. We find enhanced MFEs, up to a factor of $5$, are realized for several frequencies in the $1$--$100\,$MHz band and $J_{0}=0$, that generally outperforms the static scenario with EED interaction included but is less sensitive than the static case neglecting both interactions. However, for $J_{0}=-104\,$MHz our results demonstrate that the driven model with both exchange and EED interaction included, can be more magnetosensitive than the static case that neglects inter-radical interactions. Furthermore, when the EED interaction is included, which reflects what is realized in practice, the driven model outperforms the static case for the large range of $J_{0}$ considered and for all driving frequencies $\nu_{\mathrm{d}} \gtrsim 1\,$MHz.

\textit{Discussion.---}
Inter-radical interactions in the radical pair mechanism suppress magnetosensitivity, hindering the effectiveness of a quantum compass. We have demonstrated that a time-dependent driving, which alters the inter-radical distance, can overcome this suppression. In general, an enhancement in compass sensitivity is observed at physically plausible driving frequencies of $1$--$100\,$MHz. In nature, these driving processes could be a result of internal motion, such as low-frequency protein breathing modes and structural rearrangements following initial charge transfer or associated with sensory transduction \cite{Nordmann2017}. Sufficient amplitudes of this motion, for our identified MFE, have been predicted by molecular dynamics simulations \cite{Kattnig2018}. 

Our results suggest that processes in the natural environment of a biological system may be crucial in eliciting quantum effects. Such persistent and dynamically controllable live quantum effects, which were hypothesised by Briegel and Popescu in the context of entanglement \cite{Cai2010a, Mohseni2014}, are observed here in the context of magnetoreception. In this picture, fluctuations can create driving processes which maintain the system in a ``live'' far from equilibrium state, in contrast to their ``dead'', static counterparts \cite{Zwanzig2001}. This description of live quantum biology may provide increased robustness and explain why \textit{in vitro} experiments on isolated proteins have so far failed to reproduce the exquisite sensitivity inferred from ethological observations of live animals \cite{Hore2016}.

Our findings provide more insight into the underlying nature of magnetoreception by suggesting that the effect comprises non-adiabatic transitions driven by protein motion. This goes beyond the simple picture of hyperfine induced singlet-triplet coherences perturbed by the geomagnetic field, by highlighting the role of the protein environment \cite{Mittermaier2006} in actively supporting compass action, in contrast to it's previously understood role in solely the dissipation of quantum effects by inducing spin relaxation. Consequently, driving contributions may constitute a crucial addition to studies suggesting a beneficial stochastic environment interaction \cite{Mohseni2008, Rebentrost2009,Caruso2009, Huelga2013, Ludemann2015, Kattnig2016b,Cao2020}. To further elucidate driving in magnetoreception, we are currently pursuing more realistic models of motion, such as system-bath models with under-damped Brownian characteristics \cite{Breuer2007,Suess2014, Breuer2015, DeVega2017, Tanimura2020, Gribben2022}. Our initial finding using optimal quantum control suggests scope for further enhancing sensitivity of the natural system. Future work considering an interplay of the natural process and artificially controlled driving may provide principles for quantum inspired bio-engineering, enhancements in quantum sensing, and experimental tests of the proposed driven radical pair mechanism of magnetoreception.

\noindent We acknowledge use of University of Exeter's HPC facility. This work was supported by the UK Defence Science and Technology Laboratory (DSTLX-1000139168), the Office of Naval Research (ONR award number N62909-21-1-2018) and the pdfRC (grant EP/V047175/1). We endorse Scientific CO$_2$nduct \cite{conduct}, a CO$_2$ emission table is provided in the supplemental material.

\clearpage
\onecolumngrid 
\hoffset = 23pt
\voffset = 17pt
\textheight=649pt
\textwidth= 461pt
\fontsize{10}{14}\selectfont 
\setcounter{figure}{0} \newpage
\setcounter{equation}{0}
\renewcommand{\theequation}{S\arabic{equation}}
\renewcommand{\thefigure}{S\arabic{figure}}
\begin{center}
\textbf{\large Supplemental Material:\\ \emph{Driven} spin dynamics enhances cryptochrome magnetoreception}
\end{center}

In this supporting material, we present further justification and analysis of the claims made in the main text. Firstly, the computational details for our models are outlined. Subsequently, we provide some detail and exposition for the Floquet approach of simulating time-evolution. This is followed by analysis of coherence and entanglement in the single hyperfine-coupled nitrogen atom model. We then consider this model under further simplifications of the zero perpendicular hyperfine coupling component, and time averaged recombination and exchange, and show that general features persist for a simple two-level system. Next we consider the effect of damped driving oscillations and a variation of oscillation orientation and amplitude. Lastly, we demonstrate enhancements for a simple radical pair model driven externally with optimal quantum control, which exceeds the enhancements due to harmonic driving shown in our work. 

\section{Computational and model details}

The following open source packages were used: NumPy \cite{numpy}; SciPy \cite{scipy}; QuTiP \cite{qutip}; Matplotlib \cite{matplotlib}.

\begin{table}[h]\centering
		\begin{tabular}{lc}
		    \hline
			\textbf{Nuclei} \ & Hyperfine interaction tensor (mT) \\
			\hline
			\hline
			\\
			N5 & 
			$\begin{bmatrix}
			-0.0995 &  0.0029 & 0 \\
			0.0029 & -0.0875 & 0 \\
			0 & 0 & 1.7569 
			\end{bmatrix}$  \\
			\\
			\hline
			\\
			N10 & 
			$\begin{bmatrix}
			-0.0149 & 0.0021 & 0 \\
			0.0021 & -0.0237 & 0 \\
			0 & 0 & 0.6046 
			\end{bmatrix}$\\
			\\
			\hline
			\hline
		\end{tabular} \label{Tab.flavin}
	\caption{\label{tab:table_flavin}%
        Hyperfine coupling parameters for nuclei in the flavin radical in mT. 
	}
\end{table}
\vspace{-1.29em}
\begin{table}[h]\centering
		\begin{tabular}{lc}
		    \hline
			\textbf{Nuclei} \ & Hyperfine interaction tensor (mT) \\
			\hline
			\hline
			\\
			N1 & 
			$\begin{bmatrix}
            -0.0337 & 0.0924 & -0.1353 \\
            0.0924 & 0.3303 & -0.5318 \\
            -0.1353 & -0.5318 & 0.6679
			\end{bmatrix}$  \\
			\\
			\hline
			\\
			H1 & 
			$\begin{bmatrix}
            -0.9921 & -0.2091 & -0.2003 \\
            -0.2091 & -0.2631 & 0.2803 \\
            -0.2003 & 0.2803 & -0.5398
			\end{bmatrix}$\\
			\\
			\hline
			\hline
		\end{tabular} \label{Tab.Trp}
	\caption{\label{tab:table_tryptophan}%
    Hyperfine coupling parameters for nuclei in the tryptophan radical in mT. 
	}
\end{table}

For the main text we use hyperfine-coupling parameters\footnote{For the single hyperfine interaction model.} corresponding to the nitrogen atom N5 in flavin and additionally N10 in flavin, and N1 and H1 for tryptophan used in the more complex model study. The values for these parameters in mT are shown in Table \ref{tab:table_flavin} for flavin and Table \ref{tab:table_tryptophan} for tryptophan. Furthermore, for all simulations we assume a geomagnetic field strength of $B_{0} = 50\,\mu$T, reaction rate constants $k_{b_{0}} = 2\,\mu$s and $k_{f} = 1\,\mu$s, range parameter $\beta = 1.4\,$\AA$^{-1}$. For simulation of relative anisotropy we compute dynamics up to a maximum time of $t=12.5\,\mu$s for which the reaction dynamics are complete. Position of the tryptophan relative to flavin is given by [8.51, -14.25, 6.55] \AA. The hyperfine interactions of electron spins $\hat{\mathrm{\mathbf{S}}}_{i}$ in each radical $i$ with $n_{i}$ nuclear spins $\hat{\mathrm{\mathbf{I}}}_{i,j}$, for $j\leq n_{i}$,  is described by the hyperfine interaction Hamiltonian 
\begin{align}
    \hat{H}_{\mathrm{hf}} = \sum_{i=1}^{m}\hat{\mathrm{\mathbf{S}}}_{i} \cdot \Bigg( \sum_{j=1}^{n_{i}} \mathrm{\mathbf{A}}_{i,j} \cdot \hat{\mathrm{\mathbf{I}}}_{i,j} \Bigg),
\end{align}
where $\mathrm{\mathbf{A}}_{i,j}$ is the hyperfine coupling tensor for nuclear spin $j$ in radical $i$. The Zeeman interaction of the electron spins with the external magnetic field $\mathrm{\mathbf{B}}$ is described by $\hat{H}_{\mathrm{Zee}} = -\frac{1}{\hbar}g_{i} \mu_{B} \hat{\mathrm{\mathbf{S}}}_{i}\cdot \mathrm{\mathbf{B}}$, where $\mu_{B}$ is the Bohr magneton and $g_{i}$ is the g-value of radical $i$. The exchange interaction of the electron spins is described by $\hat{H}_{\mathrm{ex}}(t) = -2J(t)\hat{\mathrm{\mathbf{S}}}_{i}\cdot \hat{\mathrm{\mathbf{S}}}_{j}$, where the strength $J(t)$ is dependent on the inter-radical distance modulation described in the main text. 
The magnetic electron-electron dipolar interactions under the point dipole approximation are accounted for with the Hamiltonian 

\begin{align}
    \hat{H}_{\mathrm{dip}}(t) = - \sum_{i>j}^{m} D_{ij}(\vert r_{ij}(t) \vert) \Big(3(\hat{\mathrm{\mathbf{S}}}_{i}\cdot \mathrm{\mathbf{u}}_{ij})(\hat{\mathrm{\mathbf{S}}}_{j}\cdot\mathrm{\mathbf{u}}_{ij})-(\hat{\mathrm{\mathbf{S}}}_{i}\cdot \hat{\mathrm{\mathbf{S}}}_{j}) \Big),
\end{align}
where $D_{ij}(\mathrm{\mathbf{r}}_{ij}(t)) = \mu_{0} g_{i} g_{j} \mu_{B}^{2}/(4\pi \hbar \vert \mathrm{\mathbf{r}}_{ij}(t) \vert^{3})$, for which $\mu_{0}$ is the permeability of free space and $\mathrm{\mathbf{r}}_{ij}(t)$ defines the time-dependent displacement between radicals such that $\mathrm{\mathbf{u}}_{ij} = \mathrm{\mathbf{r}}_{ij}(t)/ \vert \mathrm{\mathbf{r}}_{ij}(t) \vert$ is the unit vector
\section{Floquet theory}

Floquet theory is well suited to the study of strongly driven periodic quantum systems. By essentially transforming the problem of solving a Hamiltonian with a complex time dependence (potentially involving many Fourier components) into a time-independent problem \cite{floq0}, the approach lends itself to both increased computational efficiency and scalability. One way it allows for this is by intrinsically evading secular terms (terms not periodic in the time variable \cite{floq1}). Variants of Floquet theory have seen uses in fields ranging from solid state nuclear magnetic resonance to multiphoton spectroscopy, but its application to radical pair recombination reactions under time-dependent magnetic fields was first shown in Ref.\ \onlinecite{floqKat}. 

The theory is frequently applied for quantum dynamics under unitary evolution, but also generalises for the non-unitary evolution \cite{hanggi98, ode08} induced by $H_{eff}$. Specifically, the propagator can be expressed as
\begin{align}
    U(t,0) = \phi(t) e^{-iEt}V^{-1}, \label{floq}
\end{align}
where $\phi(t)$ is the matrix of Floquet modes and the diagonal matrix of quasienergies $E$ and $V = \phi(0)$ are obtained from the eigen-decomposition of the single-period propagator
\begin{align}
    U(T, 0) = Ve^{-iEt}V^{-1}
\end{align}
The Floquet modes are periodic in $t$, i.e.\ $\phi(t) = \phi(t+T)$, and can be obtained from $\phi(t) = U(t,0)Ve^{iEt}$. So the propagator can be constructed for arbitrarily large times by integrating Eq.~(\ref{floq}) over simply two periods: once to evaluate the one-period propagator, from which $\phi(0)$ and the quasienergies are obtained, and once more subsequently to evaluate $\phi(t)$ at the required times. Due to the periodicity of $\phi(t)$, the latter only necessitates the evaluation for $t \textrm{ mod } T \in [0,T]$.


\section{Analysis of coherence and entanglement} 
Several different measures of entanglement and coherence were used to probe the radical pair mechanism in previous studies. Here, we outline a selection of these additional measures to provide further analysis of coherence and entanglement in our driven radical pair model. We first address the results of coherence and entanglement measures, which are typically defined with respect to a normalised density matrix $\hat{\rho}$ and are in general basis dependent. For instance, the relative entropy of coherence utilised in the main text,
\begin{align}
\mathcal{C}_{r}[\hat{\rho}]=S[\mathbb{I}\mathbb{C}(\hat{\rho})] - S[\hat{\rho}],   
\end{align}
is defined with respect to a basis $\{\vert n \rangle \}_{n=1}^{d}$ of the $d$-dimensional Hilbert space. Here, $S[\hat{\rho}]= - \mathrm{Tr}[\hat{\rho}\log(\hat{\rho})]$ represents the von Neumann entropy and  $\mathbb{IC}(\hat{\rho}) = \sum_{n} \vert n \rangle \langle n \vert \hat{\rho} \vert n \rangle \langle n \vert$ is a dephasing operation that returns a density matrix with its off-diagonal terms eliminated. Another commonly used coherence measure, the $l_{1}$-norm, is defined as
\begin{align}
	\mathcal{C}_{l_{1}}[\hat{\rho}] = \sum_{nm} \left\vert \langle n \vert \hat{\rho} - \mathbb{IC}(\hat{\rho}) \vert m \rangle \right\vert =  \sum_{n\neq m} \vert \langle n \vert \hat{\rho} \vert m \rangle \vert,
\end{align}
and simply sums over the moduli of the off-diagonal terms in a given density matrix. Furthermore, Kominis defined a coherence measure \cite{Kominis2020S} based on the relative entropy 
\begin{align}
	\mathcal{C}_{st}[\hat{\rho}] = S[\hat{P}_{S}\hat{\rho}\hat{P}_{S}+\hat{P}_{T}\hat{\rho}\hat{P}_{T}] - S[\hat{\rho}],
\end{align}
which alters the dephasing operation in the relative entropy of coherence to singlet and triplet projection operators $\hat{P}_{S}$ and $\hat{P}_{T}$, respectively. This measure aims to report singlet-triplet coherence while being independent of basis and unaffected by the coherence between triplet states. A similar measure of singlet-triplet coherence was used in Ref.\ \onlinecite{Kritsotakis2014S}, but is instead based on the $l_{1}$-norm measure. 

Additionally, the relative entropy of coherence  with respect to the maximally mixed state is defined as
\begin{align}
\mathcal{C}_{1} = S\left[ {\hat \rho ||\hat 1/d} \right] = \log d - S\left[ {\hat \rho } \right],    \label{SEq.basis-indep_coherence}
\end{align}
to quantify the amount of basis-independent coherence \cite{Le2020S}. These measures are often used with respect to the electronic subspace of the full density matrix $\hat{\rho}$ by performing a trace over the nuclear spin degrees of freedom $\mathrm{Tr}_{\mathrm{nuc}}(\hat{\rho}) = \hat{\sigma}$, where $\hat{\sigma}$ represents the density matrix of electron spins. In studies concerning the electron spin coherence with basis dependent measures, it is common to choose a basis consisting of spin up/down states, or of singlet/triplet states. The former provides a measure of the electron spin correlation, whereas the latter singlet and triplet states are directly related to chemical products of the radical pair reaction. Specifically, in a system of two electrons spins, there are four states in the up-down (UD) basis $\left\vert \uparrow \uparrow \right\rangle$, $\left\vert \downarrow  \downarrow \right\rangle$, $\left\vert \uparrow\mathrel{\mspace{-1mu}}\downarrow \right\rangle$ and $\left\vert \downarrow\mathrel{\mspace{-1mu}}\uparrow \right\rangle$. Alternatively, the singlet-triplet (ST) basis states are denoted by the singlet state $\left\vert S \right\rangle = (\left\vert \uparrow\mathrel{\mspace{-1mu}}\downarrow \right\rangle - \left\vert \downarrow\mathrel{\mspace{-1mu}}\uparrow \right\rangle)/\sqrt{2}$, and the triplet states $\left\vert T_{+} \right \rangle = \left\vert \uparrow \uparrow \right\rangle$, $\left\vert T_{-} \right \rangle = \left\vert \downarrow \downarrow \right\rangle$ and $\left\vert T_{0} \right\rangle = (\left\vert \uparrow\mathrel{\mspace{-1mu}}\downarrow \right\rangle + \left\vert \downarrow\mathrel{\mspace{-1mu}}\uparrow \right\rangle)/\sqrt{2} $. We make use of both basis representations to elucidate specific features with respect to driving. 

As the density matrix of the radical pair system is time-dependent, so too are measures of coherence thus far discussed. However, to analyze coherence as the exchange interaction $J_{0}$ and driving frequency $\nu_{\mathrm{d}}$ are altered, we integrate the measure over a time period relevant to the radical pair reaction. For a measure $\mathcal{C}[\hat{\sigma}(t)]$ the time-integrated coherence is given by 
\begin{align}
    \mathcal{C} = \int_{0}^{\infty} \mathcal{C}[\hat{\sigma}(t)] \mathrm{Tr}[\hat{\sigma}(t)] \label{SEq.time-int_coherence} \mathrm{dt}, 
\end{align}
which has been weighted according to the population remaining in the system using $\mathrm{Tr}[\hat{\sigma}(t)]$. In practical realisations we fix the upper limit of the time interval to $5\,\mu$s, corresponding to the time required for reaction dynamics and coherent interconversion to be complete within our model. Considerations of coherence in the radical pair mechanism also depend on the orientation dependence with respect to the magnetic field. To address this we take an average of the coherence measure
\begin{align}
    \overline{\mathcal{C}} = \frac{[\mathcal{C}_{\parallel} + \mathcal{C}_{\perp}]}{2}, \label{SEq.avg_mes}
\end{align}
where $\mathcal{C}_{\parallel}$ and $\mathcal{C}_{\perp}$ are the time-integrated coherences of the system with respect to parallel and perpendicular orientation to the magnetic field, respectively. To evaluate $\mathcal{C}_{\parallel}$ and $\mathcal{C}_{\perp}$, the time-integrated coherence of Eq.~(\ref{SEq.time-int_coherence}) must be calculated for the density matrix evolution with respect to the parallel $\hat{\sigma}_{\parallel}(t)$ and perpendicular $\hat{\sigma}_{\perp}(t)$ orientations. This measure thus reports on the average amount of coherence that is present in the system under different orientations with respect to a magnetic field.

\subsection{Average electron spin coherence with respect to orientation}

\noindent It was shown in the main text that driving which enhances sensitivity, as measured by relative anisotropy $\chi$, also stimulates coherence as measured by $\overline{\mathcal{C}_{r}}$ in the ST basis. We consider the same driven system (a single hyperfine-coupled nitrogen atom) in presence of exchange interaction with oscillation amplitude fixed at $\Delta_{\mathrm{d}} = 3\,$\AA. The single non-zero hyperfine interaction is assumed axial, with principal components given by $A_{xx} = A_{yy} = A_{\perp} = -2.6\,$MHz and $A_{zz} = A_{\parallel} = 49.2\,$MHz. The driving likewise gives rise to a time dependent recombination ${k_b}(t) = {k_{b0}}\exp \left[ { - \beta (r(t) - r_0)} \right]$ and exchange $J(t) = {J_0}\exp \left[ { - \beta (r(t) - r_0)} \right]$ with $r(t) = \frac{{{\Delta _d}}}{2}\left[ {1 - \cos (2\pi {\nu _d}t)} \right] + r_0$, with $k_{b_{0}}$ = 2\,$\mu{s}^{-1}$, $k_{f}$ = 1\,$\mu{s}^{-1}$ and $\beta = 1.4\,$\AA$^{-1}$. 

\begin{figure*}[h]
\center
\includegraphics{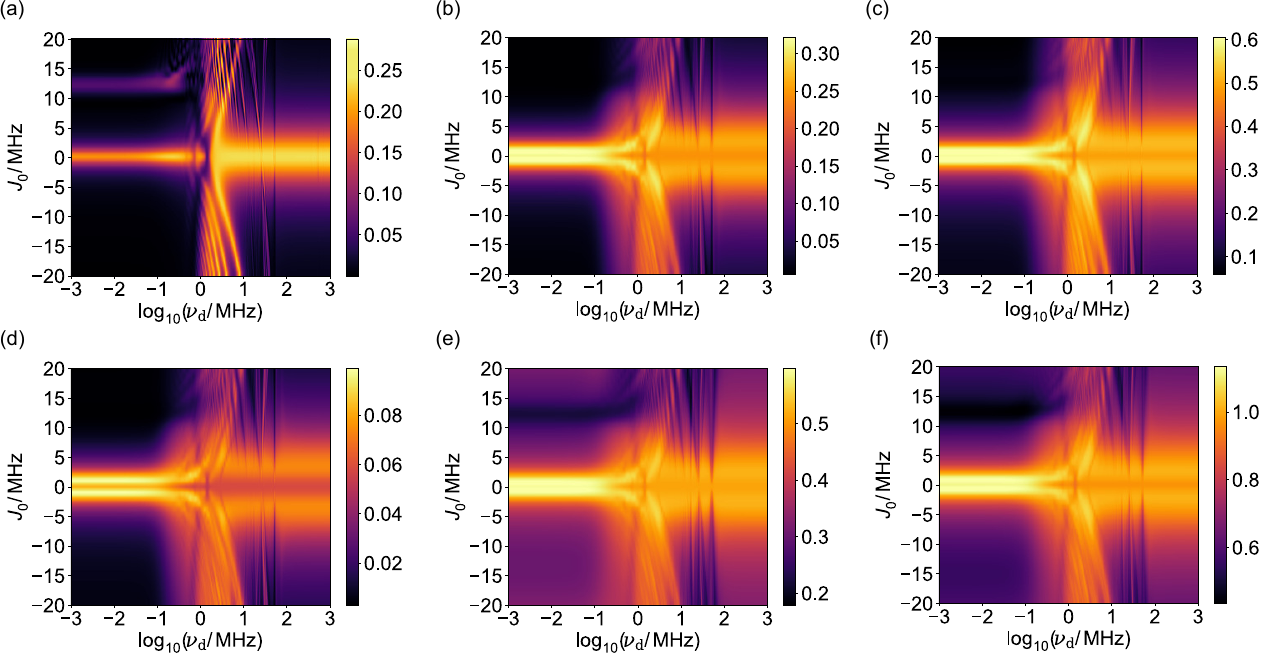}
\caption{Color maps of anisotropy and average of the electronic coherence measures with respect to orientation for a variation of exchange $J_{0}$ and driving frequency $\nu_{\mathrm{d}}$ in a model driven radical pair system comprising a single hyperfine-coupled nitrogen atom. (a) Relative anisotropy $\chi$. (b) Relative entropy of coherence $\overline{\mathcal{C}_{r}}$ evaluated in the ST basis. (c) $l_{1}$-norm of coherence $\overline{\mathcal{C}_{l_{1}}}$ evaluated in the ST basis. (d) Measure of singlet-triplet coherence $\overline{\mathcal{C}_{st}}$ (e) Relative entropy of coherence $\overline{\mathcal{C}_{r}}$ evaluated in the UD basis. (f) $l_{1}$-norm of coherence $\overline{\mathcal{C}_{l_{1}}}$ evaluated in the UD basis.  \label{SFig.cmes}}
\vspace{-2em}
\end{figure*}

In Fig.~\ref{SFig.cmes} we present color maps for the larger set of coherence measures introduced in this section, with a variation of $J_{0}$ against $\nu_{\mathrm{d}}$, and evaluate measures in both the UD and ST basis where appropriate. The relative anisotropy $\chi$ has also been included in Fig.~\ref{SFig.cmes}(a), and $\overline{\mathcal{C}_{r}}$ evaluated in the ST basis in Fig.~\ref{SFig.cmes}(b), for ease of comparison. Furthermore, we present the results of $\overline{\mathcal{C}_{l_{1}}}$ in the ST basis and $\overline{\mathcal{C}_{st}}$ in Fig.~\ref{SFig.cmes}(c \& d) which in general show qualitatively similar features to $\overline{\mathcal{C}_{r}}$. 

Larger differences are found for $\overline{\mathcal{C}_{st}}$, but it should be noted that this measure corresponds to ST coherence only, whereas the other measures are of coherence in the ST basis and also account for coherence between triplet states. Therefore, a comparison of these measures provides insight into the nature of the coherence. In general, $\overline{\mathcal{C}_{st}}$ suggests that enhancements in $\chi$ are accompanied by ST coherence,  whereas minor differences in $\overline{\mathcal{C}_{r}}$ and $\overline{\mathcal{C}_{l_{1}}}$ suggest that some coherence between triplet states is also present. 

Further insight is gained by analysing coherence as measured by $\overline{\mathcal{C}_{r}}$ and $\overline{\mathcal{C}_{l_{1}}}$ in the UD basis, as shown in Fig.~\ref{SFig.cmes}(e \& f). In particular we observe that whilst qualitative features emerge that are in agreement with ST basis measures, there is a relatively greater amount of UD coherence for low driving frequencies. By comparing UD and ST basis states, we conclude that the coherence for this parameter range is most likely due to coherence between $\left\vert \uparrow\mathrel{\mspace{-1mu}}\downarrow \right\rangle$ and $\left\vert \downarrow\mathrel{\mspace{-1mu}}\uparrow \right\rangle$ states, suggesting the system is trapped in the $\vert S \rangle$ state. However, an increase in $\chi$ in the approximate range of $10 \lesssim J_{0} \lesssim 15\,$MHz suggests that some interconversion of singlet and triplet populations occur in this range. This provides an interpretation that the interconversion is small or does not persist over time due to the lack of ST coherence and coherence between triplet states, whilst the lack of UD coherence indicates population transfer to $\vert T_{+} \rangle$ or $\vert T_{-} \rangle$ states.

\vspace{-1.4em}
\subsection{Average electron spin entanglement with respect to orientation}
\begin{figure*}[h]
\center
\includegraphics{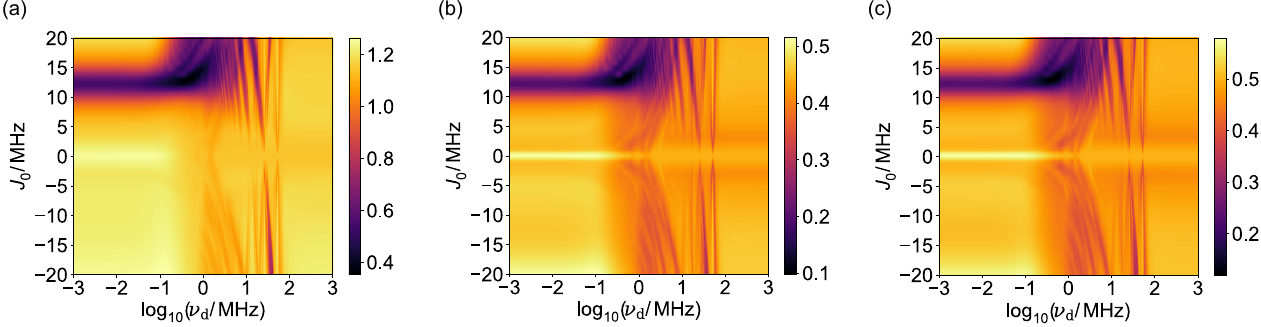}
\caption{Color maps of the average of electronic measures of entanglement and basis-independent coherence with respect to orientation for a variation of exchange $J_{0}$ against the driving frequency $\nu_{\mathrm{d}}$ in a driven radical pair system comprising a single hyperfine-coupled nitrogen atom. (a) Basis-independent coherence $\mathcal{C}_{1}$  (b) Entanglement concurrence $\overline{E_{C}}$ (c) Logarithmic negativity entanglement $\overline{E_{N}}$. \label{SFig.emes}}
\vspace{-4em}
\end{figure*}

\noindent To analyze these features further we employ entanglement measures, such as the logarithmic negativity
\begin{align}
    E_{N}[\hat{\rho}] = \log_{2} \vert \vert \hat{\rho}^{\Gamma_{A}} \vert \vert_{1},
\end{align}
where $\hat{\rho}^{\Gamma_{A}}$ is the partial transpose of $\hat{\rho}$ with respect to a subsystem $A$ and $\vert \vert \hat{A} \vert \vert_{1} = \mathrm{Tr}\vert \hat{A} \vert=\mathrm{Tr}\sqrt{\hat{A^{\dagger}}\hat{A}}$ represents the trace norm of an operator $\hat{A}$. Additionally, we compute the entanglement concurrence 
\begin{align}
    E_{C}[\hat{\rho}] = \max(0, \lambda_{1} - \lambda_{2} - \lambda_{3} - \lambda_{4}),
\end{align}
where $\lambda_{i}$ are the eigenvalues in decreasing order of 
\begin{align}
   R = \sqrt{\sqrt{\hat{\rho}} \tilde{\rho} \sqrt{\hat{\rho}}},
\end{align}
in which $\tilde{\rho} = (\sigma_{y} \otimes \sigma_{y}) \rho^{*} (\sigma_{y} \otimes \sigma_{y})$ and $\sigma_{y}$ is a Pauli spin matrix. As in the case of the coherence measures discussed, we evaluate the entanglement measures for the electron spin density matrix $\hat{\sigma}(t)$ and integrate over time as in Eq.~(\ref{SEq.time-int_coherence}). Likewise, as in Eq.~(\ref{SEq.avg_mes}), we take an average over measures evaluated with respect to the system with perpendicular and parallel orientations to the magnetic field to give 
\begin{align}
    \overline{E} = \frac{[E_{\parallel} + E_{\perp}]}{2}.
\end{align}
Note that unlike most of the coherence measures introduced, entanglement measures are independent of choice of basis. Thus, we have grouped together the basis-independent coherence of Eq.~(\ref{SEq.basis-indep_coherence}) and these entanglement measures and presented color maps of the results in Fig.~\ref{SFig.emes} for a variation of $J_{0}$ against $\nu_{\mathrm{d}}$. 
\clearpage
Qualitatively similar features emerge in all of these measures, with generally large entanglement/basis-independent coherence present across the range of parameter choices. For, low driving frequencies and large $J_{0}$, this agrees well with our interpretation that the system is trapped in the $\vert S \rangle$ state, as it corresponds to a maximally entangled state. Other parameter choices that give rise to large entanglement, may also involve the maximally entangled triplet state $\vert T_{0} \rangle$. Enhancements in $\chi$ for driving frequencies in the range $0\lesssim \nu_{\mathrm{d}} \lesssim 10\,$MHz are generally accompanied by a slight reduction in entanglement suggesting that it is also important for the system to evolve to $\vert T_{\pm} \rangle$ states. Lastly, the enhancement of $\chi$ at low driving frequencies in the approximate range of $10 \lesssim J_{0} \lesssim 15\,$MHz is accompanied by a large relative decrease in the entanglement. This further supports our interpretation based on coherence, that the system evolution for this parameter range involves a significant population transfer to either the $\vert T_{+} \rangle$ or $\vert T_{-} \rangle$ state, whilst maintaining relatively lower population in the $\vert S \rangle$ and $\vert T_{0} \rangle$ states. 
\vspace{-3em}
\subsection{Difference of electron spin coherence with respect to orientation}
\begin{figure*}[h]
\vspace{-1em}
\center
\includegraphics{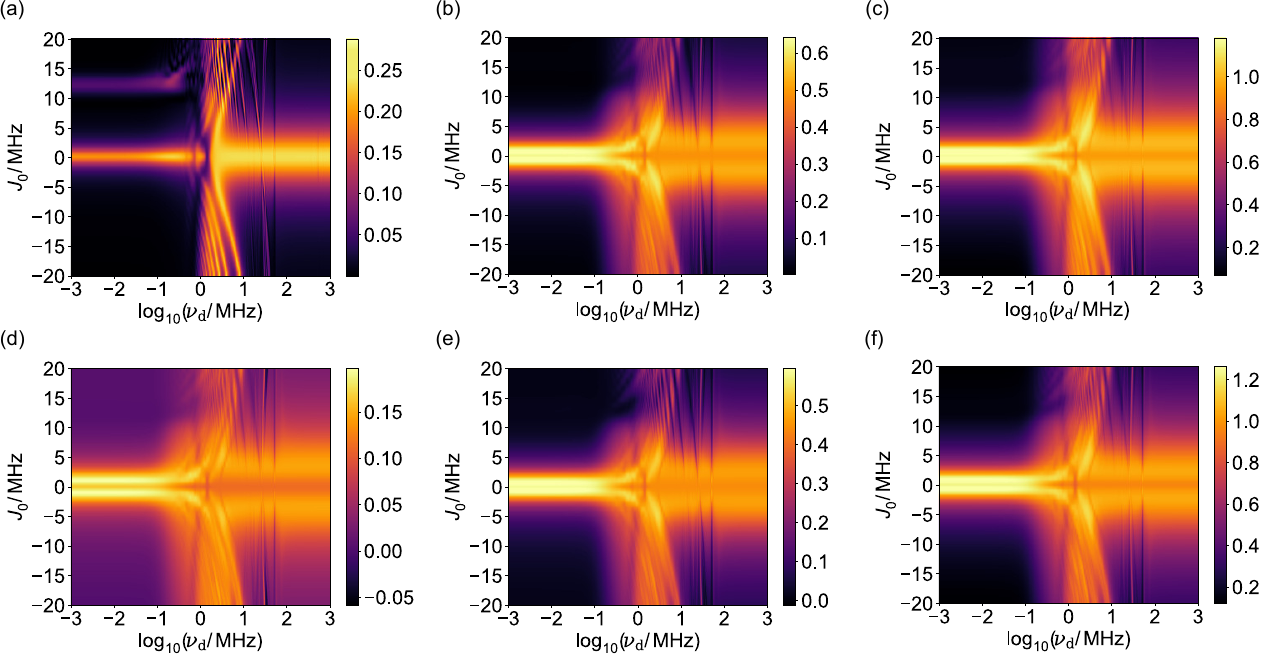}
\caption{Color maps of anisotropy and difference of the electronic coherence measures with respect to orientation for a variation of exchange $J_{0}$ and driving frequency $\nu_{\mathrm{d}}$ in a model driven radical pair system comprising a single hyperfine-coupled nitrogen atom. (a) Relative anisotropy $\chi$. (b) Relative entropy of coherence $\overline{\mathcal{C}_{r}}$ evaluated in the ST basis. (c) $l_{1}$-norm of coherence $\overline{\mathcal{C}_{l_{1}}}$ evaluated in the ST basis. (d) Measure of singlet-triplet coherence $\overline{\mathcal{C}_{st}}$ (e) Relative entropy of coherence $\overline{\mathcal{C}_{r}}$ evaluated in the UD basis. (f) $l_{1}$-norm of coherence $\overline{\mathcal{C}_{l_{1}}}$ evaluated in the UD basis.  \label{SFig.delta_cmes}}
\vspace{-2em}
\end{figure*}

\noindent In addition to taking the average of measures over parallel and perpendicular orientations of the magnetic field as in Eq.~(\ref{SEq.avg_mes}), we have defined the difference of coherence as $\Delta \mathcal{C} = \mathcal{C}_{\perp} - \mathcal{C}_{\parallel}$ and the difference of entanglement as $\Delta E = E_{\perp} - E_{\parallel}$ with respect to parallel and perpendicular orientations. This treatment aims to identify if there is an associated increase in coherence or entanglement for a given orientation and is similar in its form to the definition of anisotropy. In Fig.~\ref{SFig.delta_cmes} we plot color maps of the difference of coherence for a variation of $J_{0}$ against $\nu_{\mathrm{d}}$, using the same measures that were used for evaluating $\overline{\mathcal{C}}$. Features of the color maps largely coincide, suggesting that for parameters associated with enhancements, larger coherence is produced for the perpendicular orientation. It is also shown that there is relatively small difference in the UD coherence at low driving frequencies and large $J_{0}$ in contrast to the average of the UD coherence for this parameter range. This confirms  the system is trapped in the $\vert S \rangle$ state and is freed by specific driving frequencies that also stimulates coherence predominantly in the $\perp$ system orientation. 

\subsection{Difference of electron spin entanglement with respect to orientation}
\begin{figure*}[h]
\center
\vspace{-1em}
\includegraphics{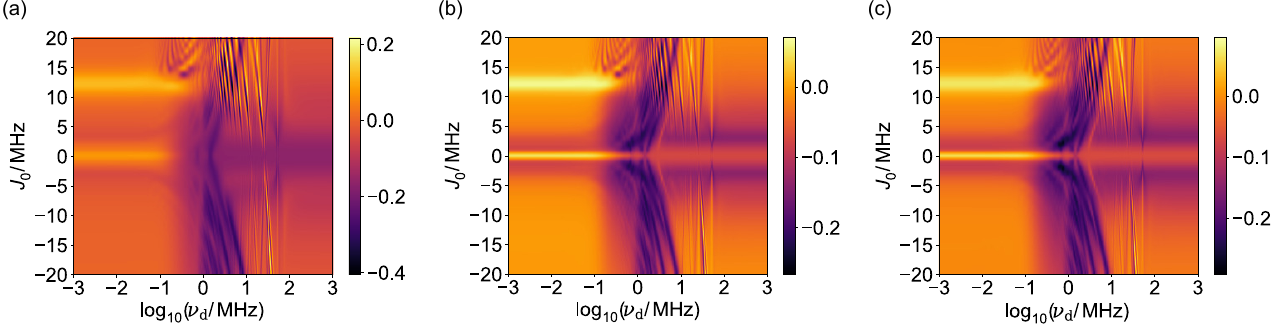}
\caption{Color maps of the difference of electronic measures of entanglement and basis-independent coherence with respect to orientation for a variation of exchange $J_{0}$ against the driving frequency $\nu_{\mathrm{d}}$ in a driven radical pair system comprising a single hyperfine-coupled nitrogen atom. (a) Basis-independent coherence $\mathcal{C}_{1}$  (b) Entanglement concurrence $\overline{E_{C}}$ (c) Logarithmic negativity entanglement $\overline{E_{N}}$. \label{SFig.diff_emes}}
\vspace{-1em}
\end{figure*}

\noindent Results of difference of coherence for the basis-independent measure of Eq.~(\ref{SEq.basis-indep_coherence}) is shown in Fig.~\ref{SFig.diff_emes}(a), whilst difference of entanglement is shown in Fig.~\ref{SFig.diff_emes}(b \& c). As was the case for the average, similar features are seen amongst different measures. In general, for low driving frequencies, the difference is close to zero, indicating the system is in the $\vert S \rangle$ state for both orientations. However, for $10 \lesssim J_{0} \lesssim 15\,$MHz a small change is observed with larger entanglement in the perpendicular orientation, suggesting that reduction in entanglement in the parallel orientation is due to redistribution of population to the $\vert T_{\pm} \rangle$ states.  
Lastly, we observe that for parameter ranges associated with enhancements the entanglement in the perpendicular orientation is lower than in the parallel. This supports our coherence motivated interpretation that some redistribution to $\vert T_{\pm} \rangle $ accompanies the enhancement in the range of $0 \lesssim \nu_{\mathrm{d}} \lesssim 10\,$MHz. 
\subsection{Electron and nuclear spin global coherence}
\begin{figure*}[h]
\vspace{-1em}
\center
\includegraphics{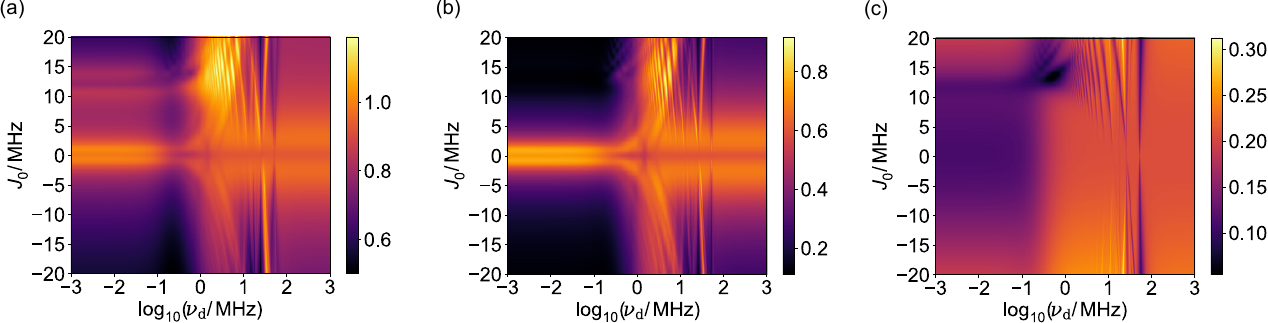}
\caption{Color maps of the global coherence for a variation of exchange $J_{0}$ against the driving frequency $\nu_{\mathrm{d}}$ in a model driven radical pair system comprising a single hyperfine-coupled nitrogen atom. (a) The average of global coherence using $\mathcal{C}_{l_{1}}$ with respect to orientations (b) The difference of global coherence using $\mathcal{C}_{l_{1}}$ with respect to orientations. (c) Global coherence measured by the coherent contribution to singlet yield $[{C}_{y}^{\mathcal{G}}]_{B=0}$. \label{SFig.glomes}}
\vspace{-1em}
\end{figure*}

\noindent In addition to the electron spin measures of coherence and entanglement we have considered, it is possible to compute a measure of the global coherence of the combined electron and nuclear system, i.e, of $\hat{\rho}$. We have computed this for the $l_{1}$ norm of coherence, as before time-integrating as in Eq.~(\ref{SEq.time-int_coherence}), and taking the average or difference over orientations with results shown in Fig.~\ref{SFig.glomes}(a \& b). In Fig.~\ref{SFig.glomes}(c) we show a global measure of coherence defined by Cai and Plenio \cite{Cai2013S} which constitutes the field-independent ($B=0$) singlet recombination yield due to the coherent part of the initial density operator $\mathbb{GC}(\hat{\rho}(0))$, with $\mathbb{GC}(\hat{\rho}) = \hat{\rho} - \mathbb{IC}(\hat{\rho})$ evaluated in the eigenbasis of the hyperfine Hamiltonian:
	\begin{align}
		[\mathcal{C}_{y}^{\mathcal{G}}]_{B=0} = \Big\vert Y_{S}\Big(\hat{\rho}(0) = (z_{1}z_{2})^{-1}\mathbb{GC}(\hat{P}_{S});B=0\Big)\Big\vert.
	\end{align}

In general the global coherence features do not coincide as strongly with the relative anisotropy $\chi$ enhancements for the simple model considered. However, certain parameter ranges such as $10 \lesssim J_{0} \lesssim 20\,$MHz and $0 \lesssim \nu_{\mathrm{d}} \lesssim 10\,$MHz, suggest that global coherence as measured by $\mathcal{C}_{l_{1}}$ may play a role in the spin dynamics of specific enhancements. In contrast, features of global coherence as measured by $[\mathcal{C}_{y}^{\mathcal{G}}]_{B=0}$ are less intricate and show the global coherence increases for driving frequencies above $\nu_{\mathrm{d}}\approx 1\,$MHz, with some further increases observed at large $J_{0}$.

\subsection{Time-resolved coherence and entanglement}
\noindent We have thus far time-integrated the coherence and entanglement to analyze different parameter choices of $J_{0}$ and $\nu_{\mathrm{d}}$. Here, to understand the dynamical properties of driving on entanglement and coherence we analyze the measures as a function of time. Live quantum effects have previously been considered in the context of dynamical entanglement by Cai \textit{et al} \cite{Cai2010aS}, who demonstrated that entanglement can persistently recur in a simple two-spin system that is periodically driven and coupled to a noisy environment. 

\begin{figure*}[h]
\center
\includegraphics{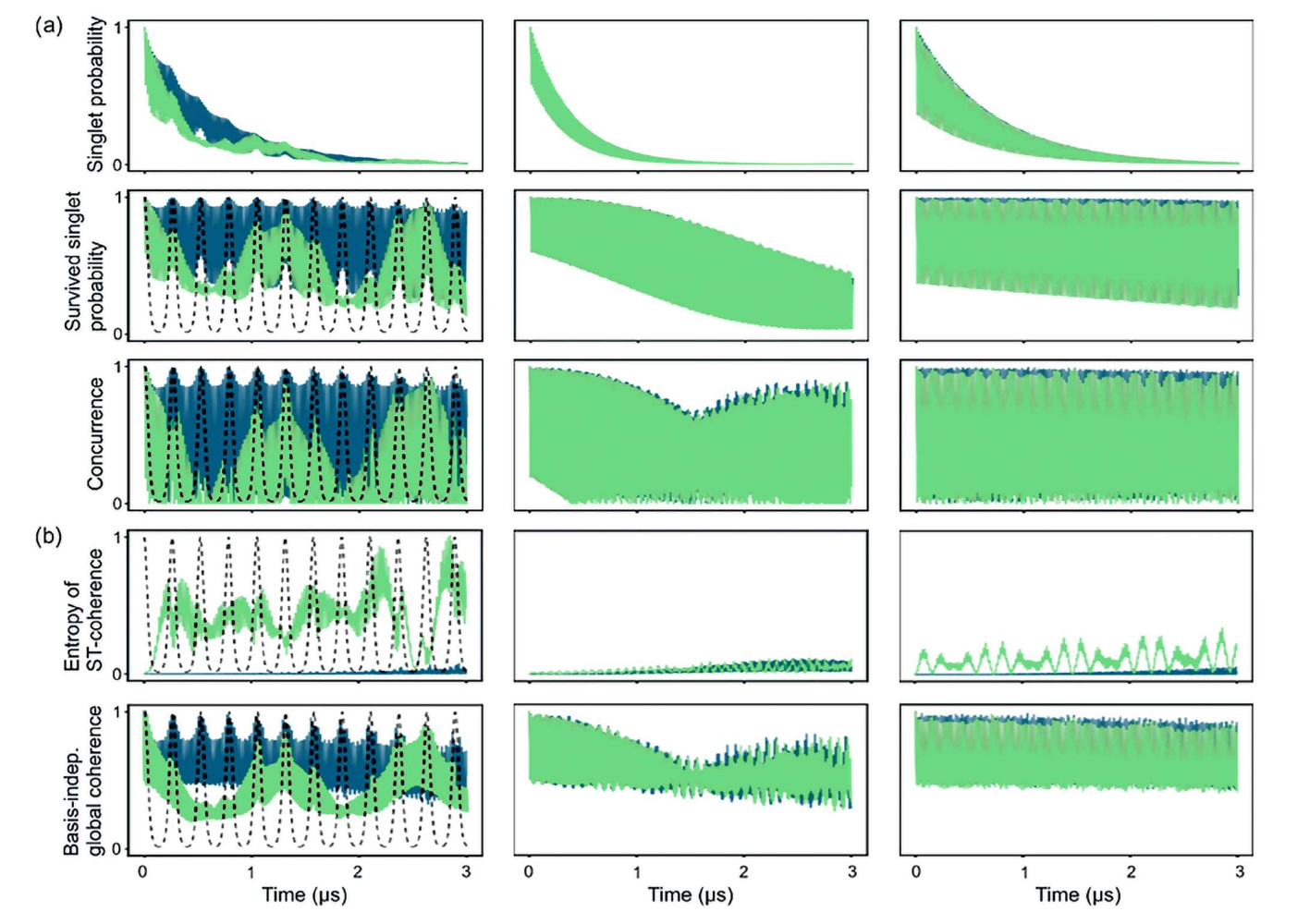}
\caption{In (a), we show singlet probability, survived singlet probability, and entanglement concurrency for a one-nitrogen radical pair. Results are shown for a parallel (dark blue) or perpendicular (bright green) orientation of the principal hyperfine axis with respect to the magnetic field with a fixed driving oscillation amplitude $\Delta_{\mathrm{d}} = 3\,$\AA, exchange interaction strength $J_{0}=20\,$MHz, and driving frequency $\nu_{\mathrm{d}}=3.8\,$MHz. For the set of simulations displayed on the left, the exchange coupling and singlet recombination yield were time-dependent as described in our driving model. For the central column and right column of data these two parameters were constant and equal to their respective value at $t=0$ or their time-average over one driving period $1/\nu_{\mathrm{d}}$, respectively. For the driven system, the dashed black lines illustrate the time-dependence of the exchange coupling or recombination rate constant normalised to $1$ for ease of comparison. In (b), we present the relative entropy of coherence $\mathcal{C}_{r}$ evaluated in the ST basis, and basis-independent coherence of the one-nitrogen radical pair. For the central and right columns of data the exchange and recombination rate were constant and equal to their respective value at $t=0$ or their time-average over one driving period $1/\nu_{\mathrm{d}}$, respectively. For the driven system, the dashed black lines illustrate the time-dependence of the exchange coupling or recombination rate constant normalised to $1$ for ease of comparison \label{SFig.time-res}}
\vspace{-2em}
\end{figure*}
\begin{figure*}[b]
\center
\includegraphics[]{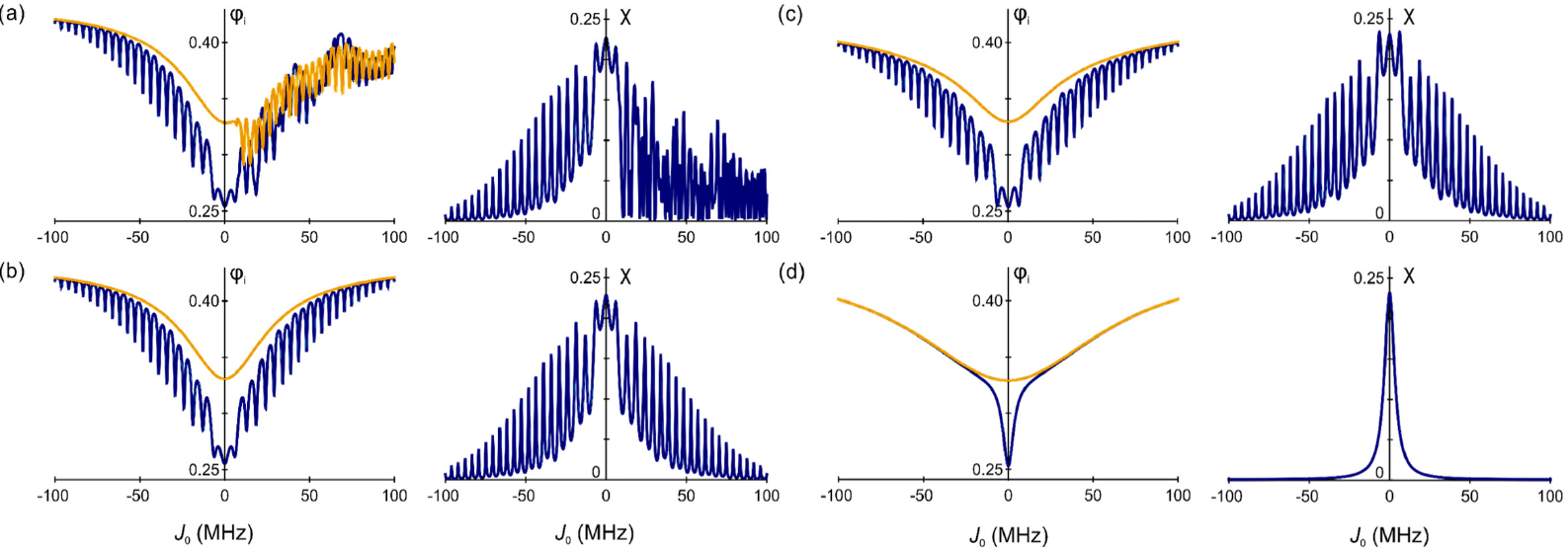}
\caption{Recombination yields and associated compass fidelities for a one-nitrogen radical pair with axial hyperfine interaction (principal component $A_{\parallel} = 49.2\,$MHz and scalar inter-radical coupling, ${\hat H_{12}} =  -2J\,{{\bf{\hat S}}_1} \cdot {{\bf{\hat S}}_2}$). For (a) with $A_{\perp} = -$2.6 MHz, $J(t)$ and the singlet recombination rate $k_{b}(t)$ are time-dependent, whereas for (b)-(d) $A_{\perp} = 0$. For (c) recombination rate  $k_{b}(t)$ is replaced by the time-average over one oscillation period. Finally, for d) the time-averaged quantities were used in place of both $J(t)$ and $k_{b}(t)$, i.e. the effective Hamiltonian is time-independent. \label{SFig.Recomb_yields}}
\vspace{-1.2em}
\end{figure*}

 
\clearpage
Assuming that the spin coupling and the energy gap of the system are modulated by the classical motion, the authors have demonstrated the cyclic generation of fresh entanglement, even under conditions for which the static thermal state is separable for all possible spin-pair configurations. For the one-nitrogen radical pair model introduced in the main text, and a parameter choice of driving oscillation amplitude $\Delta_{\mathrm{d}} = 3\,$\AA, exchange interaction strength $J_{0}=20\,$MHz, and driving frequency $\nu_{\mathrm{d}}=3.8\,$MHz, we also observe that the driving motion periodically boosts the electronic entanglement. This entanglement rejuvenation is evident in the singlet probability of the survived radical pairs as well as the concurrence of the electronic spin $E_{C}[\hat{\sigma}(t)]$, as is shown in Fig.~\ref{SFig.time-res}(a). Both measures peak as the radical pairs are driven to small inter-radical distances, i.e. when the inter-radical coupling is engaged. Note however that the effect is present for both canonical orientations of the static magnetic field, suggesting that entanglement rejuvenation does not directly promote high compass fidelity. Figure~\ref{SFig.time-res}(b) shows that large compass sensitivity of the driven systems concurs with the generation of large singlet-triplet electronic coherence as assessed via $\mathcal{C}_{r}[\hat{\sigma(t)}]$ evaluated in the ST basis and a large reduction of global coherence (as assessed by a basis-independent coherence measure) for the perpendicular magnetic field direction. We further show that for zeroed perpendicular hyperfine couplings or a time-averaged recombination rate the orientation dependence is largely removed. We consider these system simplifications in more detail in the next section for the recombination yields and associated compass fidelities (see Fig.~\ref{SFig.Recomb_yields}).

\section{Further simplifications of the driven radical pair model}

Using the same inter-radical distance driving as described in the main text, we first consider the case of a single hyperfine-coupled nitrogen atom (I = 1) in one radical and no hyperfine interactions in the other. Likewise, the single non-zero hyperfine interaction has been assumed axial, with principal components given by $A_{xx} = A_{yy} = A_{\perp} = -2.6\,$MHz and $A_{zz} = A_{\parallel} = 49.2\,$MHz, and reaction rates $k_{b_{0}}$ = 2\,$\mu{s}^{-1}$, and $k_{f}$ = 1\,$\mu{s}^{-1}$. An arbitrary driving frequency of $\nu_{\mathrm{d}}= 3\,$MHz, within the identified effective range, and oscillation amplitude of $\Delta_{\mathrm{d}} = 2\,$\AA\ is chosen. For modulation of the inter-radical coupling strength $J_0$, Fig.~\ref{SFig.Recomb_yields}(a) shows recombination yields $\Phi_{\parallel}$, and $\Phi_{\perp}$ for a static magnetic field pointing in parallel and perpendicular directions, respectively, and the relative anisotropy $\chi$ defined as $\vert \Phi_{\parallel} - \Phi_{\perp} \vert / \max(\Phi_{\parallel}, \Phi_{\perp})$. We seek to reduce complexity of this model while retaining essential features, like resilience to inter-radical coupling, to further understand the nature of the effect. To that end, we first observe that the effect is not crucially dependent on the non-zero perpendicular hyperfine coupling components. While it is true that the magnetic field effect (MFE) is more complex for $ A_{\perp}\neq0$, the main features are retained for $ A_{\perp}=0$, as shown in Fig.~\ref{SFig.Recomb_yields}(b). Second, we note that the effect is not critically, i.e. as concerning qualitative features, dependent on the time-dependence of the recombination processes, which in essence corresponds to a stroboscopic observation of singlet state dynamics. Specifically, substituting the time-dependent $k_{b}(t)$ by its average over integer multiples of the driving period, we obtain results as shown in Fig.~\ref{SFig.Recomb_yields}(c), where the same qualitative picture emerges as for Fig.~\ref{SFig.Recomb_yields}(b). With respect to this, time average is calculated as 
\begin{align}
\left\langle {{e^{ - \beta \,\Delta (r(t) - r_0)}}} \right\rangle {\rm{ = }}{e^{ - \frac{{\beta \,{\Delta _d}}}{2}}}{I_0}\left( {\frac{{\beta \,{\Delta _d}}}{2}} \right), \label{eq.time_average}
\end{align}
where the range parameter $\beta = 1.4\,$\AA$^{-1}$\ and $I_0$ denotes the modified Bessel function of first order $n = 0$.

However, if we also average the time-dependence of the exchange coupling using Eq.~(\ref{eq.time_average}), the enhancements are diminished and we reproduce the static radical pair mechanism result of MFE suppression due to large inter-radical interactions (see Fig.~\ref{SFig.Recomb_yields}(d)). Thus, we conclude that the effect of driving inter-radical distance, as observed here, is predominantly due to the time-dependence of the inter-radical coupling and not dependent on the non-zero $A_\perp$ or the time-dependence of the recombination process. Note however that if the non-zero $A_\perp$ was retained, an analogous result to Fig.~\ref{SFig.Recomb_yields}(d) would have been obtained except for a second peak of magnetosensitivity at $\approx -12\,$MHz that emerges in the static limit as a consequence of a level crossing (i.e. for $A_{\perp}=0$; level-anti-crossing for $A_{\perp}\neq0$; in zeroth order, for the magnetic field parallel to $A_{\parallel}$ the crossing occurs at $J=\frac{1}{4}(2\omega_0-A_{\parallel})$between $\vert T_{+},m_I=-1 \rangle$ and $\vert S,m_I=0 \rangle$).

In the main text, the Hamiltonian for the magnetic field pointing along the perpendicular direction was shown for $A_{\perp}=0$. We now further analyze the structure of the Hamiltonian with $A_{\perp}=0$ for a static magnetic field applied in the parallel direction. As the Hamiltonian commutes with the $z$-component of the nuclear spin, $\hat{I}_{z}$, for both directions of the applied field, the magnetic quantum number of the nitrogen atom ${m_I} \in \{ 1,0, - 1\} $ can be used to label the energy eigenstates. For the field pointing along the parallel z-axis, with basis states chosen as $\left| {{T_ + }} \right\rangle $, $\left| {{T_0}} \right\rangle $, $\left| {{T_ - }} \right\rangle $ and $\left| S \right\rangle $, the ${m_I}$-sector of the Hamiltonian is given as
\begin{align}
\hat{H}_{\parallel} =
\begin{pmatrix}
am_{I} +\omega_{0} - J & 0 & 0 & 0\\
0 & -J & 0 & am_{I}\\
0 & 0 & -am_{I} - \omega_{0} - J & 0\\
0 & am_{I} & 0 & J
\end{pmatrix},
\end{align}
where $a = {A_\parallel }/2$ is introduced for succinctness. Similarly, for static magnetic field applied along the x-axis
\begin{align}
\hat{H}_{\perp} =
\begin{pmatrix}
am_{I} - J & b & 0 & 0\\
b & -J & b & am_{I}\\
0 & b & -am_{I} - J & 0\\
0 & am_{I} & 0 & J
\end{pmatrix},
\end{align}
where $b=\omega_{0}/\sqrt{2}$. We have already described how the inclusion of driving introduces a Landau-Zener type transition that mediates transitions between the $\vert S\rangle$ and $\vert T_{0}\rangle$ state. Here, by comparing $\hat{H}_{\perp}$ and $\hat{H}_{\parallel}$, it is revealed that $\hat{H}_{\perp}$ allows further evolution to $\vert T_{\pm} \rangle$ states, whereas no additional coherent interconversion with $\vert T_{\pm} \rangle$ states is possible for $\hat{H}_{\parallel}$. This contributes to the magnetosensitivity of the system by enacting a different response for perpendicular and parallel magnetic field orientations. In the case of inter-radical coupling and $A_{zz}$ that are comparable in magnitude, MFEs are small for the static system as the dynamics are characterised by fast $\vert S \rangle$-$\vert T_{0} \rangle$ interconversion. Consequently, there is only a marginal magnetic field response for $\hat{H}_{\perp}$ and $\hat{H}_{\parallel}$ is independent of the magnetic field. However, the inclusion of driving from molecular motion enhances directional magnetosensitivity as the periodic reduction of inter-radial coupling enables $\vert T_{\pm} \rangle$ redistribution for $\hat{H}_{\perp}$ (leading to reduction in singlet population), but it is unaffected for $\hat{H}_{\parallel}$.
\begin{figure*}[h]
\center
\includegraphics[scale=1.2]{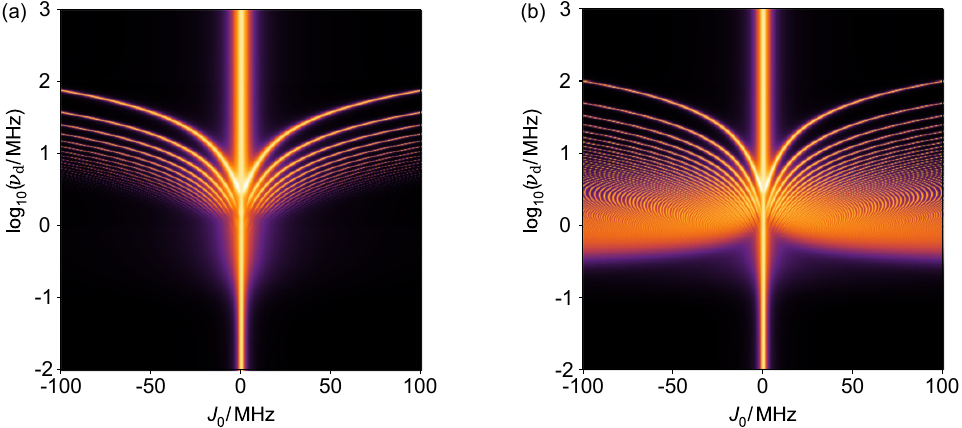}
\caption{Color map of state-transition efficiency, for two-level system with Hamiltonian $\hat H(t) = J(t){\hat \sigma _z} + b{\hat \sigma _x}$, with constant $b=1.4\,$MHz, and $J(t)$ periodic with period $1/\nu_{\mathrm{d}}$ as function of driving frequency $\nu_{\mathrm{d}}$ and the maximal energy gap $J_{0}$. For (a) $J(t) = {J_0}\exp \left[ { - \beta (r(t) - r_0)} \right]$ with  $r(t) = \frac{{{\Delta _d}}}{2}\left[ {1 - \cos (2\pi {\nu _d}t)} \right] + r_0$ while for (b) $J(t) = {J_0}\left[ {1 - (r(t) - r_0)/{\Delta _d}} \right]$, i.e. $J(0)$ is harmonically oscillating between $J_{0}$ and $0$. We chose an oscillation amplitude of $\Delta_{\mathrm{d}} = 2\,$\AA, and range parameter $\beta = 1.4\,$\AA$^{-1}$ The transition efficiency is evaluated as $\int_0^\infty  {k\exp ( - kt){{\left| {c_2(t)} \right|}^2}{\rm{dt}}} $, where $k = 1\,\mu $s$^{-1}$  and ${\left| {c_2(t)} \right|^2}$ is probability of finding the system in the initially unoccupied state. \label{SFig.two_level_system}}
\end{figure*} 

For more complex systems with several hyperfine couplings, as seen in the main text, intricate magnetic field dependent spin dynamics can arise, with non-adiabatic transitions initiated by avoided-crossings that occur at non-zero $J_0$. For such scenarios, our description provides only a qualitative picture. However, a simpler two level system with energy-gap driven as assumed above, i.e. with $\hat H(t) = J(t){\hat \sigma _z} + b{\hat \sigma _x}$ (where ${\hat \sigma _i}$ are Pauli operators), exemplifies that general features persist as complexity is increased, as is shown in Fig. (\ref{SFig.two_level_system}). Specifically, spikes in magnetosensitivity, as both driving frequencies $\nu_{\mathrm{d}}$ and $J_{0}$ are modulated, are associated with the minima and maxima of the Floquet quasienergies of the system. 

\section{Simple model under Optimal Quantum Control}
Optimal Quantum Control comprises the class of optimization algorithms which use electromagnetic ``pulses'' designed using a minimal time approach \cite{heg13, deff17} to drive a given quantum system to reach a particular target state. Without loss of generality, the Hamiltonian for a system undergoing quantum coherent spin dynamics can be written as: 
\begin{align}
\hat{H}(t) = \hat{H}_{0} + \sum_{k=1}^{n} f_{k}(t) \hat{H}_{k},
\end{align} 
with $\hat{H}_0$ being the drift Hamiltonian describing the time-independent part of the system, $f_{k}$ are the control fields and $\hat{H}_k$ is the set of control Hamiltonians coupling the fields to the system, for example via Zeeman or dipole interactions. The values taken by controls for a system undergoing quantum coherent spin dynamics, which in our case corresponds to modulation of the inter-radical distance, can be parameterised by piece-wise constant control amplitudes in the time domain. 

\begin{figure}[h]
\center
\includegraphics[scale=1.16]{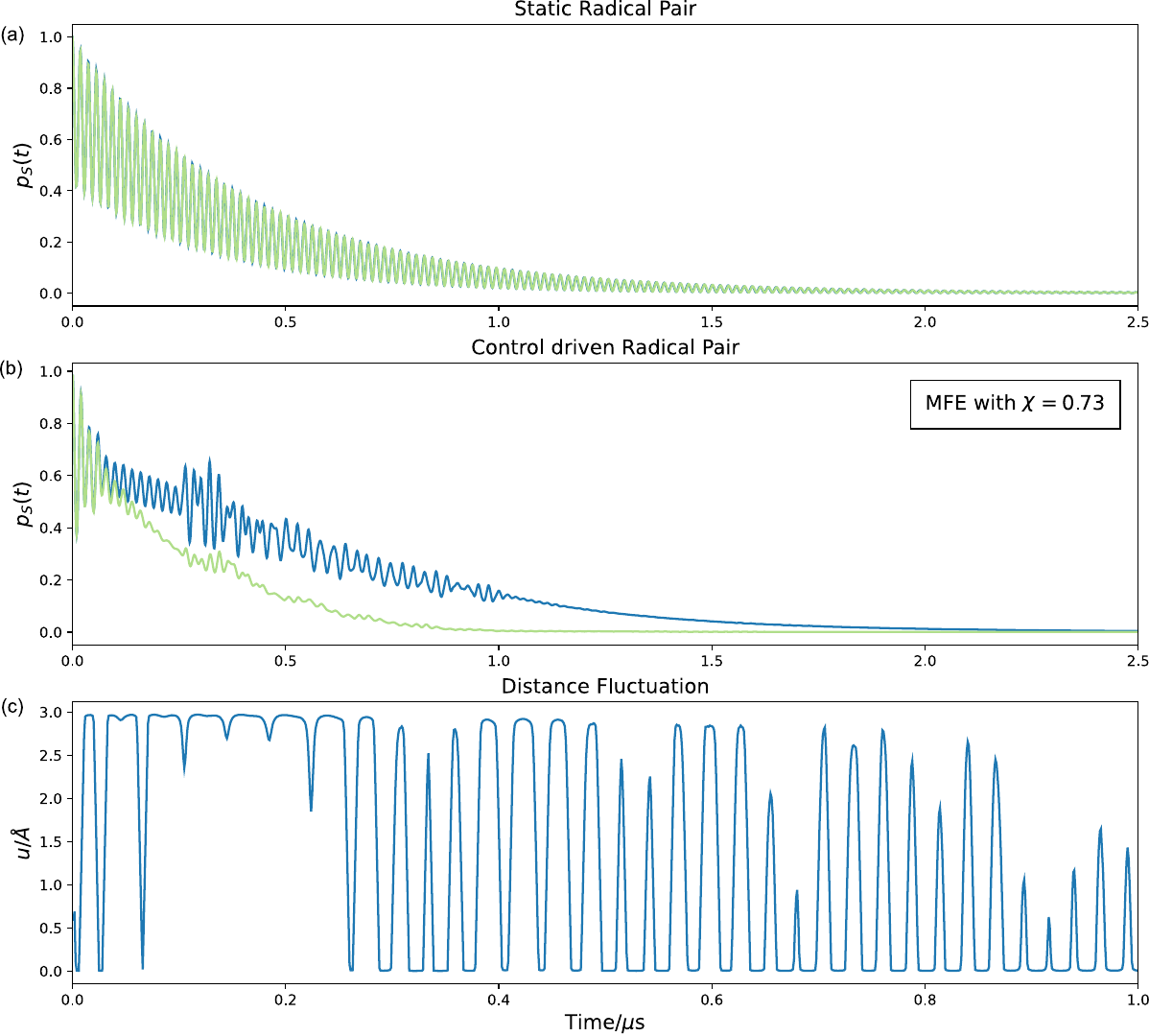}
\caption{Singlet yield population as a function of time for quantum control driven one nitrogen radical pair ($I = 1$) subject to distance-dependent exchange interaction and recombination. Given a time period of 10 nanoseconds and a roughly $100\,$MHz frequency, the maximal timescale for a transition from 3\,\AA\ to 0\,\AA\ is 1 amplitude per 5 nanoseconds. The static case is shown in (a). For control driven case (b), absolute anisotropy is minimised using optimization with 4000 time stpdf and 200 sampling points, with green corresponding to the the hyperfine tensor for the magnetic field along the axis $A_{\perp}$ and blue corresponding to the hyperfine tensor for the magnetic field along the axis $A_{\parallel}$. In (c), the distance fluctuation for the quantum control driven case is shown.}
\label{Fig.control}	
\end{figure}

We apply optimal quantum control in this sense to a single hyperfine-coupled nitrogen radical pair, with principal components given by $A_{xx} = A_{yy} = A_{\perp} = -2.6\,$\,MHz and $A_{zz} = A_{\parallel} = 49.2\,$MHz, $k_{b_{0}}$ = 2\,$\mu{s}^{-1}$, $J_0=10$\,MHz, $k_{f}$ = 1\,$\mu{s}^{-1}$. A “speed limit” of 3\,\AA\ is imposed to ensure that the inter-radical distance does not fluctuate at unrealistic speeds. 
Using pulses to externally drive spin dynamics in the radical pair, we induce hyperfine interaction mediated transitions between $|S \rangle$/$|T_{\pm}\rangle$ states. Since  we have what boils down to essentially a bilinear control problem, the optimal solution jumps between two boundaries, and so control of the \emph{bang-bang} type is optimal, with a single control function $f(t)$ and states reaching their target at fixed time intervals. By formulating a piecewise constant Hamiltonian, with total time subdivided into fixed intervals, cost functions are computed by sampling the intervals by a few time points. We optimize the escape yield, with maximal yield subject to the speed limit, using a GRAPE \cite{oqc05} based approach with standard Python optimization packages. Using this method, we obtain well optimized distance fluctuations with a large magnetic field effect, as seen in  Fig.~\ref{Fig.control}.

This demonstrates how optimal quantum control can enhance the magnetic field effect in a radical pair reaction, and the enhancement obtained by this (artificial) external driving considerably exceeds what is achievable via the harmonically driven natural systems considered elsewhere in this work.

\section{Variation of driving oscillation orientation and amplitude}
\begin{figure*}[h]
\center
\includegraphics[scale=1.1]{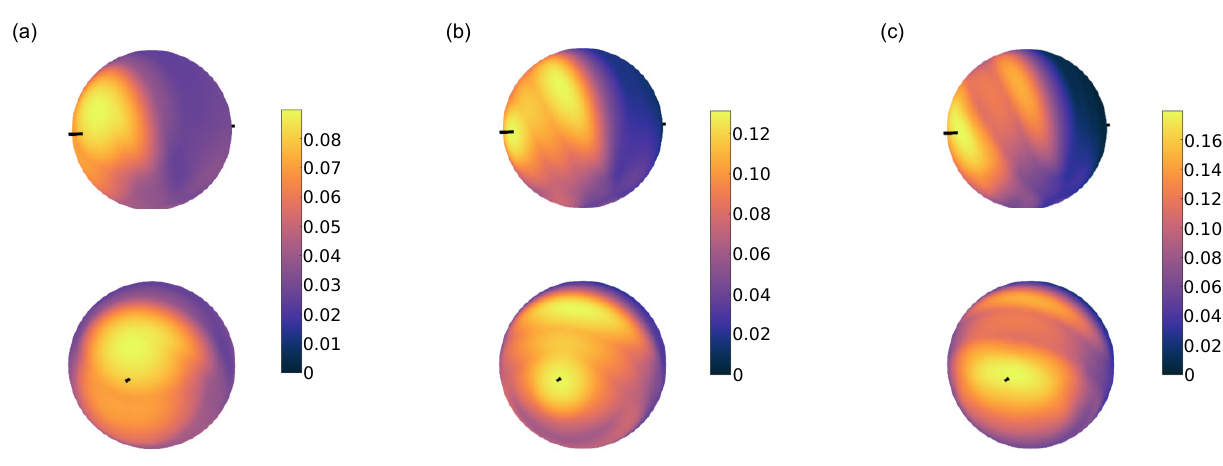}
\caption{Orientation dependence is displayed for a driven model comprising a single hyperfine-coupled nitrogen with EED interaction, and a fixed driving frequency of $\nu_{\mathrm{d}} = 4.4\,$MHz. Oscillation amplitudes of (a) $\Delta_{\mathrm{d}} = 2\,$\AA, (b) $\Delta_{\mathrm{d}} = 4\,$\AA, and (c) $\Delta_{\mathrm{d}} = 6\,$\AA, indicate effectiveness for broad increases of inter-radical distance.}
\label{SFig.osc_amp}
\end{figure*}

In the main text we demonstrated the orientation dependence of driving enhancements for a single hyperfine-coupled nitrogen atom with electron-electron dipole (EED) interactions and oscillation amplitude $\Delta_{\mathrm{d}}=2\,$\AA. In Fig.~\ref{SFig.osc_amp} we present further cases of $\Delta_{\mathrm{d}}=4\,$\AA, and $\Delta_{\mathrm{d}}=6\,$\AA, which also show oscillations that broadly increase the inter-radical distance are effective without necessitating  exact alignment. However, inter-radical modulation is more effective if it has a closer alignment with the inter-radical axis, an observation that becomes more pronounced as oscillation amplitude is increased. 

\begin{figure*}[h]
\center
\includegraphics[scale=1.1]{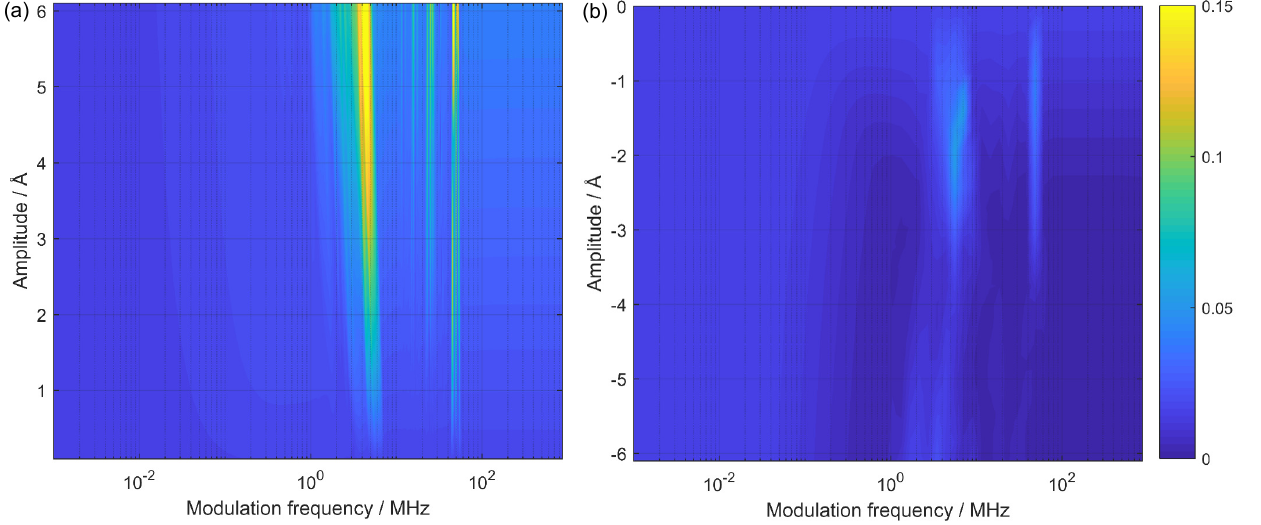}
\caption{Driven radical pair model with EED interactions included and $J_{0}=0$. Color map of relative anisotropy $\chi$ is shown for a variation of oscillation amplitude $\Delta_{\mathrm{d}}$ against driving frequency $\nu_{\mathrm{d}}$. Driving that increases the inter-radical distance is shown on the left whilst driving that decreases it is shown on the right.
\label{SFig.osc_red}}
\vspace{-1.8em}
\end{figure*}

In Fig.~\ref{SFig.osc_red} we also consider driving that reduces the inter-radical distance rather than our standard approach of increasing it periodically. Although enhancements are more prominent for driving that increases the distance, the effect is seen to persist qualitatively for a decrease of the distance, suggesting that the time-dependent nature of the driving is also of importance. 

\section{Damped driving oscillations}
In the natural setting driving may be hindered by constrained protein motion or interaction with the environment leading to damped oscillations. Here, we analyze this by considering damped inter-radical distance modulation $r(t) = r_{0} + \frac{\Delta_{\mathrm{d}}}{2}[1-\cos(2\pi \nu_{\mathrm{d}}t)] \exp(-\tau^{-1} t)$, with recombination rate $k_{b}(t) = k_{b_{0}}\exp \left[ { - \beta [r(t)-r_{0}} \right]]  $, and  $J(t) = {J_0}\exp \left[ { - \beta [r(t)-r_{0}} \right]]$, for a choice of, $k_{b_{0}}=2\,\mu$s, range parameter $\beta = 1.4\,$\AA$^{-1}$, and oscillation amplitude of $\Delta_{\mathrm{d}}=1.5\,$\AA. We study the case of the single hyperfine-coupled nitrogen atom with exchange interaction introduced in the main text. 
\begin{figure*}[h]
\center
\includegraphics[scale=1.2]{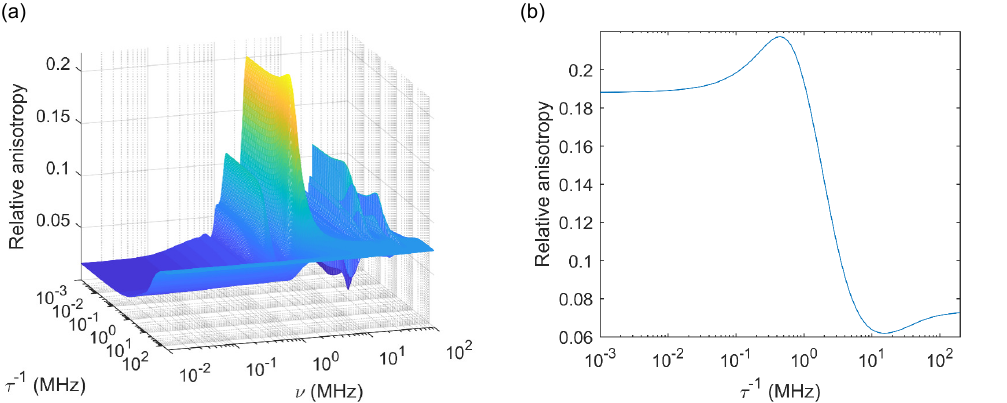}
\caption{Relative anisotropy $\chi$  for a choice of $J_{0}=-5\,$MHz, range parameter $\beta = 1.4\,$\AA$^{-1}$, and oscillation amplitude of $\Delta_{\mathrm{d}}=1.5\,$\AA. (a) Shown as a function of the inverse of the damping period $\tau^{-1}$ and driving frequency $\nu_{\mathrm{d}}$. (b) Fixed driving frequency at maximum $\chi$ as a function of $\tau^{-1}$. \label{SFig.damp_osc}}
\vspace{-1.2em}
\end{figure*} 

The relative anisotropy for this model is shown in Fig.~\ref{SFig.damp_osc} as a function of driving frequency $\nu_{\mathrm{d}}$ and inverse of the damping period $\tau^{-1}$. The results indicate that for oscillations that are damped too swiftly, i.e. for $\tau^{-1}\gtrsim1\,$MHz, there is a relative decrease in the anisotropy $\chi$ as the system will begin to resemble the case of static radical pairs. For weaker damping with $\tau^{-1}\lesssim 0.1\,$MHz, the modulation of inter-radical distance persists for longer and resembles the damping free regime studied in the main text which shows relative enhancement over the static case. However, the optimal result is found for when damping and driving occur on a similar timescale to the reaction dynamics in the range of $1-10\,\mu$s.

\section{CO$_2$-emission table}
In Table \ref{tab:table_co2}, we summarize climate expenses, calculated according to guidance found in \cite{conductS}.
\begin{table}[h]
\begin{tabular}[b]{lc}
\hline
\textbf{Numerical simulations} & \\
\hline
\hline
Total Kernel Hours [$\mathrm{h}$]& 268800\\
Thermal Design Power Per Kernel [$\mathrm{W}$]& 5.9\\
Total Energy Consumption [$\mathrm{kWh}$] & 1586\\
Average Emission Of CO$_2$ [$\mathrm{kg/kWh}$]& 0.47\\
\textbf{Were the Emissions Offset?} & Yes\\
\hline
Total CO$_2$-Emission [$\mathrm{kg}$] & 745\\
\hline
\hline
\end{tabular}
    \label{tab:co2}
    \caption{\label{tab:table_co2}%
    Carbon footprint associated with the numerical simulations. }
\end{table}

The CO$_2$ emission per kWh used is the average global value. To try to offset the $745$kg emitted, a donation was made to Atmosfair (\url{https://atmosfair.de}), a not-for-profit that promotes, develops and finances renewable energies in over fifteen countries worldwide.

\end{document}